\documentclass[twocolumn]{aa}
\usepackage{graphicx,times,amssymb}
\begin{document}
   \title{Rotation periods for very low mass stars in the Pleiades}

   \author{Alexander Scholz\thanks{Visiting Astronomer, 
   German-Spanish Astronomical Centre, Calar Alto, operated by 
   the   Max-Planck-Institute for Astronomy, Heidelberg, jointly 
   with the Spanish National Commission for Astronomy}
   \and
   Jochen Eisl{\"o}ffel}

   \offprints{A. Scholz, e-mail: scholz@tls-tautenburg.de}

   \institute{Th{\"u}ringer Landessternwarte Tautenburg,
              Sternwarte 5, D-07778 Tautenburg, Germany}

   \date{Received 12/12/2003; accepted 16/03/2004}
   
   \authorrunning{A. Scholz and J. Eisl{\"o}ffel}
   
   \titlerunning{Rotation periods for VLM stars in the Pleiades}

   \abstract{We present the results of a photometric monitoring campaign
   for very low mass (VLM) members of the Pleiades. Periodic photometric 
   variability was detected for nine VLM stars with masses between 0.08 
   and 0.25\,$M_{\odot}$. These variations are most likely caused by
   co-rotating, magnetically induced spots. In comparison with solar-mass
   stars, the photometric amplitudes are very low ($<0.04$\,mag), implying
   that either the fraction of the spot-covered area, the asymmetry of the 
   spot distribution, or the contrast between spots and photospheric 
   environment decreases with mass. From our lightcurves, there is
   evidence for temporal evolution of the spot patterns on timescales
   of about two weeks. The rotation periods range from 2.9\,h to 40\,h and tend to 
   increase linearly with mass. Compared with more massive stars, we clearly 
   see a lack of slow rotators among VLM objects. The rotational evolution of 
   VLM stars is investigated by evolving the previously published periods for 
   very young objects (Scholz \& Eisl{\"o}ffel \cite{se04}) forward in time,
   and comparing them with those observed here in the Pleiades. We 
   find that the combination of spin-up by pre-main sequence contraction and
   exponential angular momentum loss through stellar winds is able to reproduce 
   the observed period distribution in the Pleiades. This result may be 
   explained as a consequence of convective, small-scale magnetic fields.
   
   }

   \maketitle

\section{Introduction}
\label{intro}

Rotation is a key parameter of stellar evolution. The investigation
of the rotational evolution of solar-mass stars showed that the
angular momentum regulation is directly connected to basic
stellar physics: Solar-mass stars rotate slowly in the T Tauri
phase, probably because of rotational braking through magnetic 
coupling between star and disk. After loosing the disk, the rotation
accelerates as the stars contract towards the zero age main 
sequence (ZAMS). From here, the rotation rates decrease again
as a consequence of angular momentum loss through stellar winds
(Bouvier et al. \cite{bfa97}, Bodenheimer \cite{b95} and
references herein; see also the recent reviews of Stassun \&
Terndrup \cite{st03} and Mathieu \cite{m03}). Main ingredients
for models of rotational evolution are thus angular momentum loss
via a) magnetic interaction between star and disk and b) 
stellar winds.

One of the cornerstones of our understanding of rotational
evolution are the rotation rates for Pleiades members, since the age of the 
Pleiades (125\,Myr, Stauffer et al. \cite{ssk98}) marks the beginning 
of the main sequence for solar-mass stars, and thus 
the turning point in their angular momentum evolution. From the work of 
Magnitskii (\cite{m87}), Stauffer et al. (\cite{ssb87}), van Leeuwen et al. 
(\cite{lam87}), Prosser et al. (\cite{pss93}, \cite{psm93}, \cite{psd95}), 
and Krishnamurthi et al. (\cite{ktp98}), we have a large number of 
photometric rotation periods for Pleiades stars with $0.5<M<1.2\,M_{\odot}$
in hand, available from the Open Cluster Database compiled by C.F. Prosser and 
J.R. Stauffer. For very low  mass (VLM) stars with $M<0.4\,M_{\odot}$, however, 
there are only two periods known (Terndrup et al. \cite{tkp99}). First insights 
into the rotational behaviour of the VLM members of the Pleiades 
come from rotational velocity studies (Terndrup et al. \cite{tsp00} and references
herein), indicating a lack of slow rotators among VLM stars. This has been
explained by a mass-dependent saturation threshold of the angular
momentum loss rate (e.g., Barnes \& Sofia \cite{bs96}, Krishnamurthi et al. 
\cite{kpb97}). 

Whereas the rotational velocity analysis suffers from projection effects and 
high uncertainties, rotation periods can be determined
from photometric lightcurves without ambiguity with respect to the 
inclination angle and with high precision. Considering the lack of known 
rotation periods for VLM objects, it is necessary to compile a period
database that complements the known periods for solar-mass stars. This was the
main motivation for our long-term project dedicated to the study of rotation
periods in the VLM regime. In the first publication of this project, we 
presented 23 rotation periods for VLM objects in the very young cluster
around $\sigma$\,Ori (Scholz \& Eisl{\"o}ffel \cite{se04}, hereafter SE2004), giving
us the first, although rough approximation for the initial period distribution 
of VLM objects. Here we report the discovery of nine new rotation
periods for VLM stars in the Pleiades, increasing the VLM period
sample for this cluster by a factor of 4.5. With the periods from the 
$\sigma$\,Ori cluster and the Pleiades, we are now able to set contraints on the
angular momentum regulation in VLM objects during the first $10^8$\,years
of their evolution.

The paper is structured as follows: In Sect. \ref{obs}, we outline the 
selection of our targets, the time series observations, the image reduction,
and the photometry. We then report about the time series analysis in 
Sect. \ref{tsa}. In Sect. \ref{ori}, we discuss the origin of the
observed photometric variability. Subsequently, in Sect. \ref{massper},
we investigate the mass dependence of the Pleiades periods, by
comparing our data with periods for solar-mass stars and $v\sin i$ 
measurements. Then, we try to reconstruct the VLM period distribution
in the Pleiades from our periods for younger objects, taking into 
account basic angular momentum regulation mechanisms in Sect. \ref{ageper}. 
Finally, we present our conclusions in Sect. \ref{conc}.

\section{Targets, observations, data reduction}
\label{obs}

The Pleiades are a preferred hunting ground for Brown Dwarfs and
VLM stars. Here, the first cluster Brown Dwarfs at all were detected 
(Rebolo et al. \cite{rzm95}, \cite{rmb96}). In the last decade, a number 
of deep, large surveys explored the mass function of this cluster well down 
into the substellar regime (e.g., Moraux et al. \cite{mbs03}, Adams et al. 
\cite{asm01}, Bouvier et al. \cite{bsm98}).  Our target sample is based 
on the survey of Pinfield et al. (\cite{phj00}, \cite{pdj03}), which 
covers six square-degrees and is complete down to $I_C=19.6$, corresponding 
to a mass of roughly 0.05\,$M_{\odot}$ (Baraffe et al. \cite{bca98}). By the 
time of our observations, this was the largest and deepest object sample
available. Pinfield et al. (\cite{phj00}) observed in the I- and
Z-band for the primary photometric identification of their candidates.
Contaminating field stars were rejected based on near-infrared
photometry and proper motions. Their cluster member list comprises
339 objects, including 30 Brown Dwarfs. This survey confirms
numerous cluster member candidates from other studies. 

\begin{figure}[htbd]
\centering
\resizebox{\hsize}{!}{\includegraphics[angle=-90,width=6.5cm]{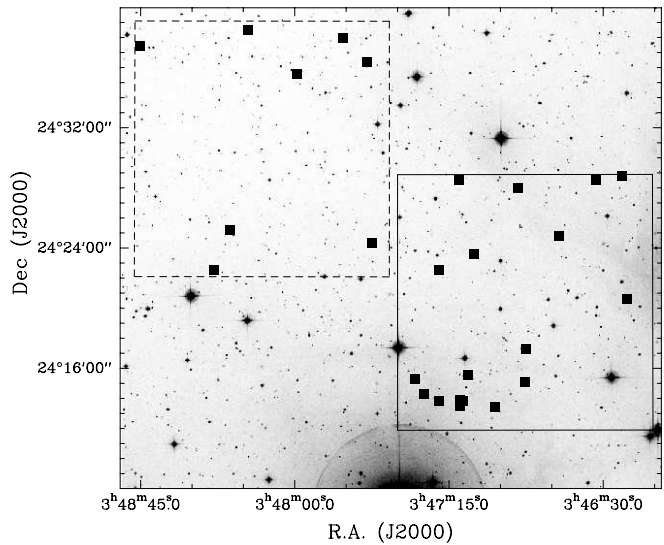}}
\caption{Time series fields in the Pleiades: The image shows our two fields
(field A - solid lines, field B - dashed lines) plotted over a DSS image. 
The squares mark our targets, which are Pleiades members from the survey
of Pinfield et al. (\cite{phj00}).}
\label{field}
\end{figure}

A subsample of the Pinfield et al. list serves as targets for
our photometric monitoring campaign. We used the CCD camera at
the 1.23-m telescope on Calar Alto, which has a field of view
of $17'\times17'$ and a pixel scale of 0$\farcs$5/pix. All time
series images were taken in the I-band and with 600\,s exposure time.
To increase the number of targets, we decided to observe two fields. 
These two fields were selected by maximising the number of cluster 
members on the detector area. Both fields together contain 39 
Pleiades members with masses up to $0.5\,M_{\odot}$. Of these, 
13 are too bright and hence saturated on our deep images. The remaining
26 targets have masses between 0.06 and 0.25$\,M_{\odot}$. The positions 
of our fields and these 26 candidates are shown in Fig. \ref{field}.

\begin{figure}[htbd]
\resizebox{\hsize}{!}{\includegraphics[angle=-90,width=6.5cm]{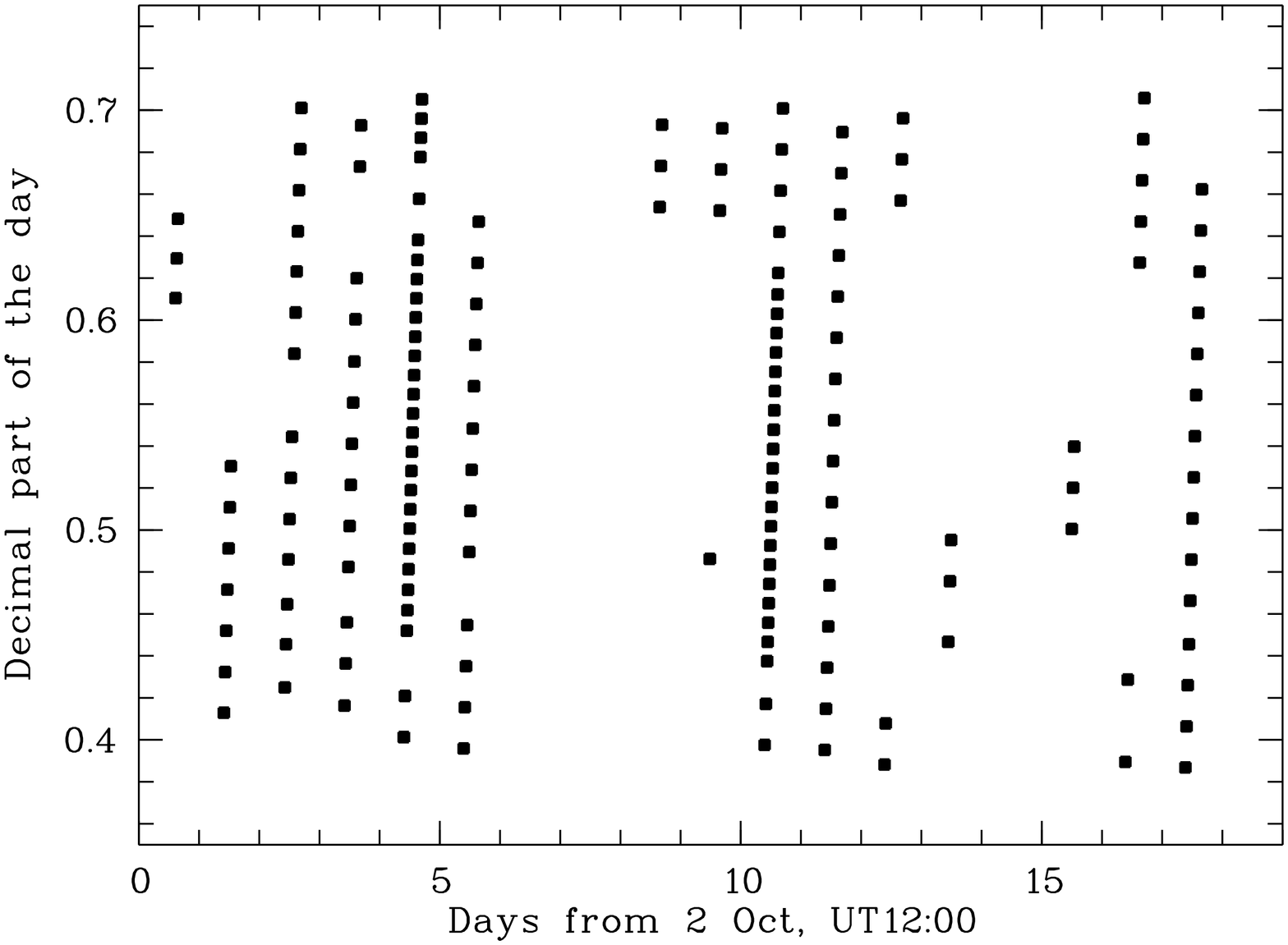}}
\resizebox{\hsize}{!}{\includegraphics[angle=-90,width=6.5cm]{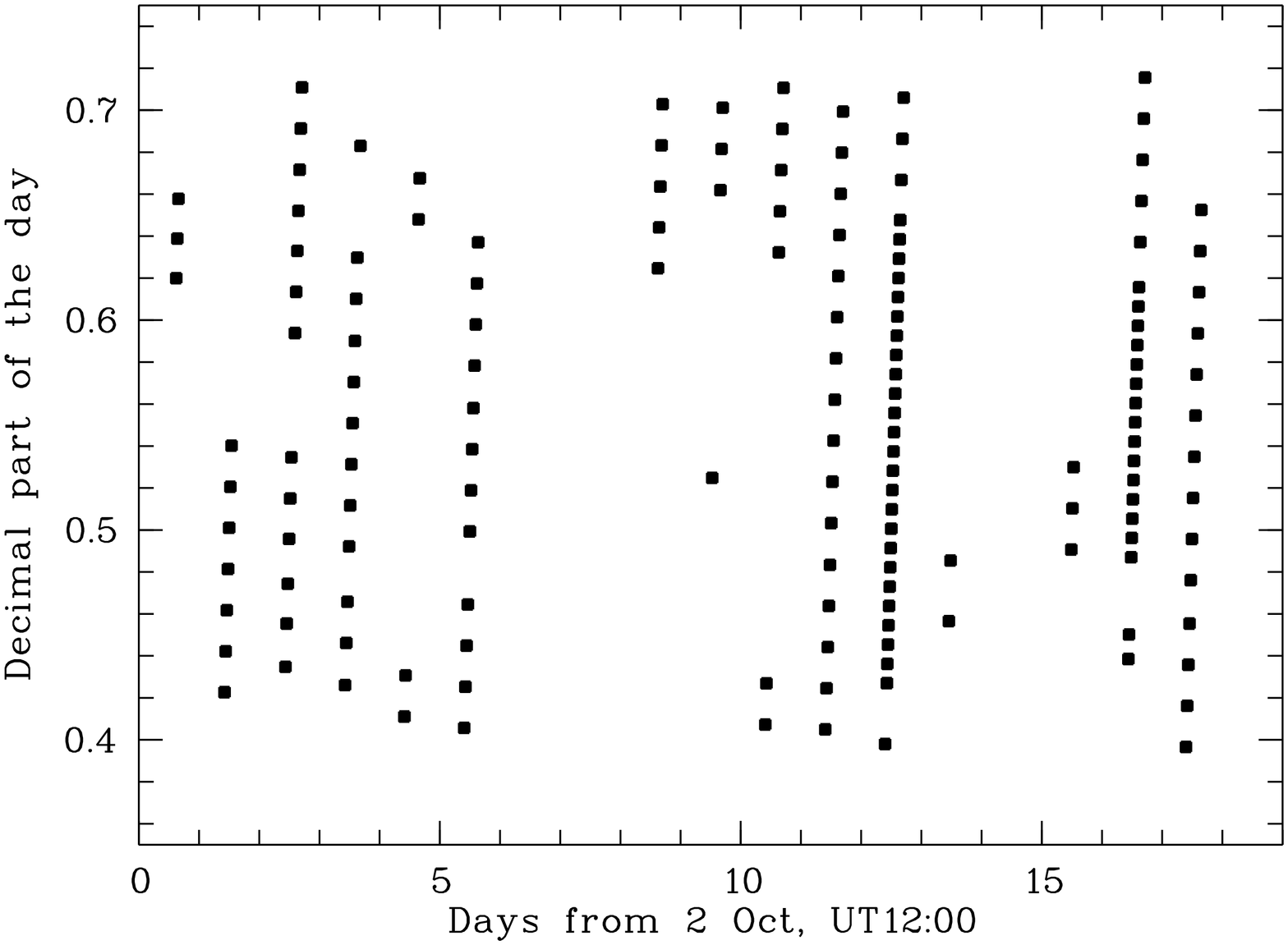}}
\caption{Observing data distribution for the time series campaign in field
A (upper panel) and field B (lower panel). Plotted is the non-integer fraction of 
the observing time against the observing time.}
\label{sampling}
\end{figure}

Our time series covers 18 nights from 2. to 19. October 2002. Observations 
were possible in 15 nights within this time span, and in seven nights we were 
able to obtain more than 10 images per field. The sampling is thus quite dense
(see Fig. \ref{sampling}). In 10 nights, both fields were monitored alternately. 
The remaining 4 observing nights were used to increase our sensitivity for very 
short periods by concentrating mainly on one field (i.e. 2 nights for each). 
Therefore, the sampling is slightly different for the two fields.

The reduction of the time series images included bias subtraction, flatfield
correction, and fringe removal (see SE2004 for details). For the further 
processing, we used the difference image analysis package of the 
'Wendelstein Calar Alto Pixellensing Project' (see Riffeser et al. \cite{rfg01},
G{\"o}ssl \& Riffeser \cite{gr02}), which is based on 'Optimal Image
Subtraction' (Alard \& Lupton \cite{al98}). This package is particularly
well-suited for CCD image from the Calar Alto 1.23-m telescope and improves
the photometric precision by several mmag compared with PSF fitting photometry,
as we have demonstrated in SE2004. 

The result of the 'Optimal Image Subtraction' pipeline are difference images which 
contain only photon noise and variable sources. On these frames, we performed 
aperture photometry for all objects. A catalogue of the pixel positions of all 
objects was determined previously using SExtractor (Bertin \& Arnouts \cite{ba96}) 
on a stacked image. By dividing the fluxes from the difference image through the 
fluxes from the stacked image, we calculated relative fluxes for all objects in 
all frames. These relative fluxes were then transformed to relative magnitudes 
with $m_{rel} = -2.5\log_{10}(1+f_\mathrm{rel})$. We determined the photometric 
precision of our photometry by calculating mean and rms of all lightcurves, after 
excluding 3$\sigma$ outliers. Similar to our result in SE2004, we achieve a 
precision of 4\,mmag for the brightest targets.

\section{Time series analysis}
\label{tsa}

For the time series analysis, we used the procedures described in SE2004. 
In a first step, the lightcurve of each target was inspected visually.
We find no signs of sudden brightness eruptions (like flares). The 
following lightcurve analysis is described in the next two subsections. 

\subsection{Generic variability test}
\label{gen}

\begin{figure}[htbd]
\centering
\resizebox{\hsize}{!}{\includegraphics[angle=-90,width=6.5cm]{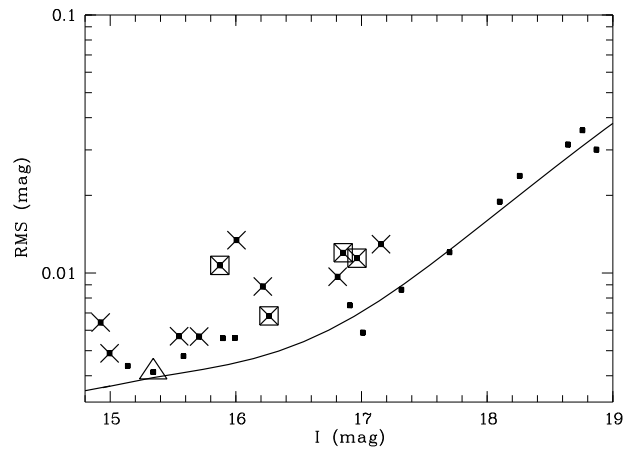}}
\caption{Lightcurve rms vs. magnitude for all targets: The solid
line shows our photometric precision, determined by fitting the rms of all
stars in our fields with a low degree polynomial. The VLM Pleiades members
are shown as dots, and all variable objects among them are marked with a cross.
The four objects, which are variable but without periodicity are additonally
marked with a square. One object with significant periodic variability, but
without variability detection in the generic test, is marked with a 
triangle (object BPL150, see Sect. \ref{per}). The magnitudes on the x-axis are 
instrumental, though the deviations from the I-band magnitudes given by Pinfield 
et al. (\cite{phj00}) are $\le0.3$\,mag.}
\label{rms}
\end{figure}

To examine the photometric variability of our VLM Pleiades members, we analysed 
the scatter in their lightcurves. To increase the sensitivity, this test was 
done after binning the lightcurves by a factor of four, i.e. we averaged the 
relative magnitudes over two hours observing time. In Fig. \ref{rms}, we plot 
the rms of all candidate lightcurves $\sigma_{cand}$ after excluding $\ge 3\sigma$ 
outliers. The solid line in Fig. \ref{rms} was determined by fitting the rms of 
the lightcurves of all stars in both fields with a low degree polynomial. Thus,
this function gives us an estimate for the average rms of the lightcurves of 
nonvariable stars ($\sigma_{av}$). Using the statistical F-test, we compared 
the rms of each candidate lightcurve $\sigma_{cand}$ with $\sigma_{av}$ for the 
respective brightness. All variable objects on the significance level of 99\% are 
marked with a cross. Out of 26 VLM members of the Pleiades, 12 (46\%) are variable 
according to this criterion.
The amplitudes of the variability are below 2\% in all cases. This result shows 
that VLM objects in the Pleiades exhibit photometric variability, but to a very low 
degree. We find no large amplitude variability, as detected for similar mass objects 
in the much younger cluster around $\sigma$\,Ori (SE2004). This is not surprising, 
since the large amplitude variations there are probably caused by accretion processes, 
which are not expected for the older objects in the Pleiades. 

For eight of the twelve variable objects, we found significant periodic 
variability (see Sect. \ref{per}). These objects will be discussed in the 
following Section. The remaining four targets clearly show anomalies in their 
lightcurves.  The lightcurves of objects BPL111 and BPL128
(catalogue number of Pinfield et al. \cite{phj00}) show noticeable variability 
only in part of the time series data. For BPL111, the mean relative magnitude deviates 
in the last two nights by $-0.015$\,mag (night 17) and $+0.03$\,mag (night 18) 
from the average of the whole time series. For object BPL128, the variation is
increased by 30\% in the second third of the lightcurve compared to the remaining
data points. The behaviour of these two objects is most likely caused by evolving 
surface properties, and will be discussed in Sect. \ref{ori}. The lightcurves
of the objects BPL130 and BPL177 exhibit several single data points which lie
$2\ldots 3\sigma$ above or below the average. Excluding these data points, the objects 
appear to be non-variable. The nature of these outliers remains unclear, and 
has to be verified by monitoring with better time resolution.

\subsection{Period search}
\label{per}

Our period search is based on the Scargle periodogram (Scargle \cite{s82}). 
It includes a sequence of tests to control the significance of a detected periodicity and
to assure that the period is intrinsic to the target. Furthermore, it uses the CLEAN 
algorithm (Roberts et al. \cite{rld87}) to distinguish between real periodogram
peaks and artifacts. The period search procedure requires that five criteria are 
fulfilled:

\begin{figure*}[htbd]
\resizebox{5.9cm}{!}{\includegraphics[angle=-90,width=6.5cm]{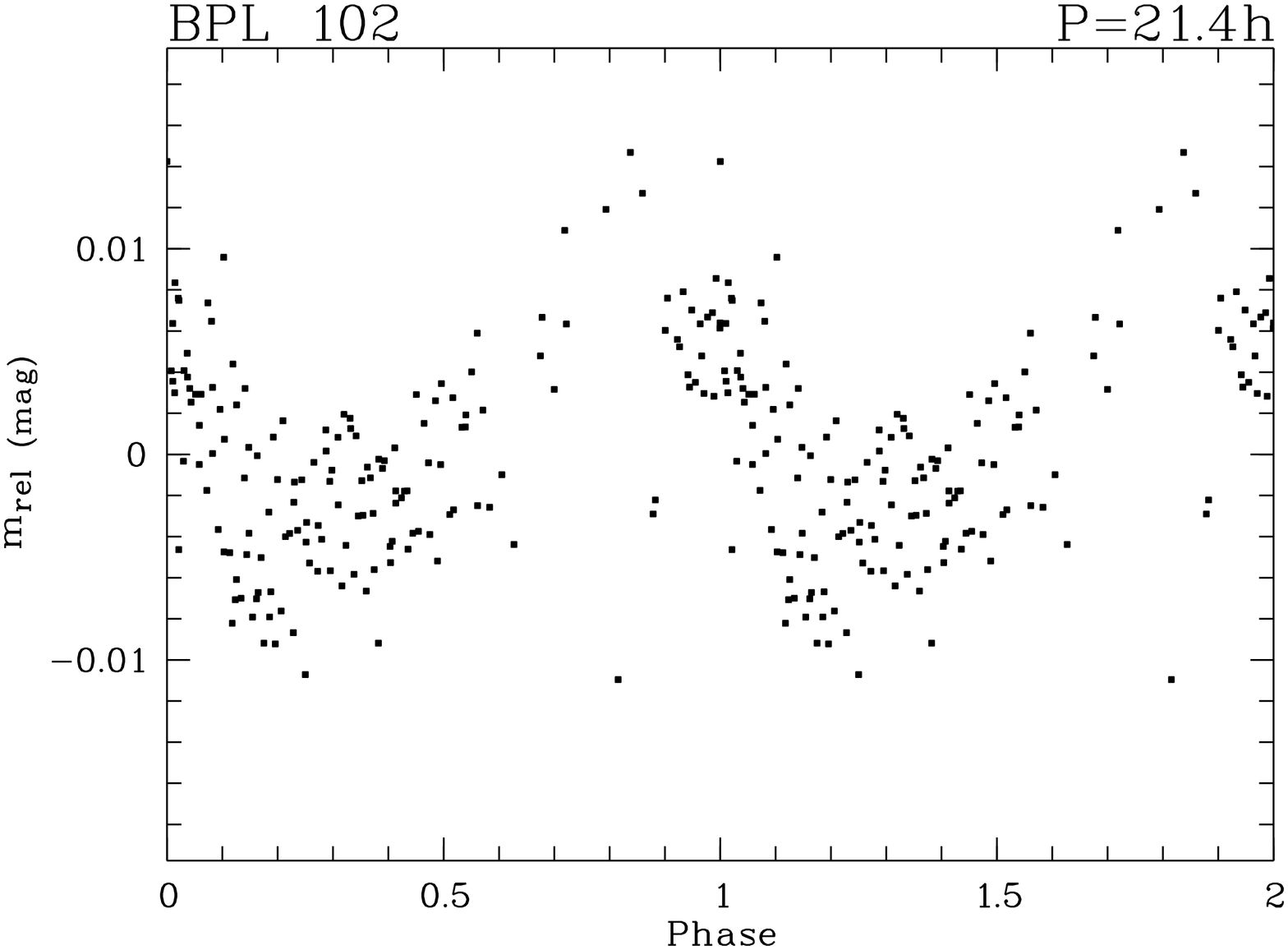}} \hfill
\resizebox{5.9cm}{!}{\includegraphics[angle=-90,width=6.5cm]{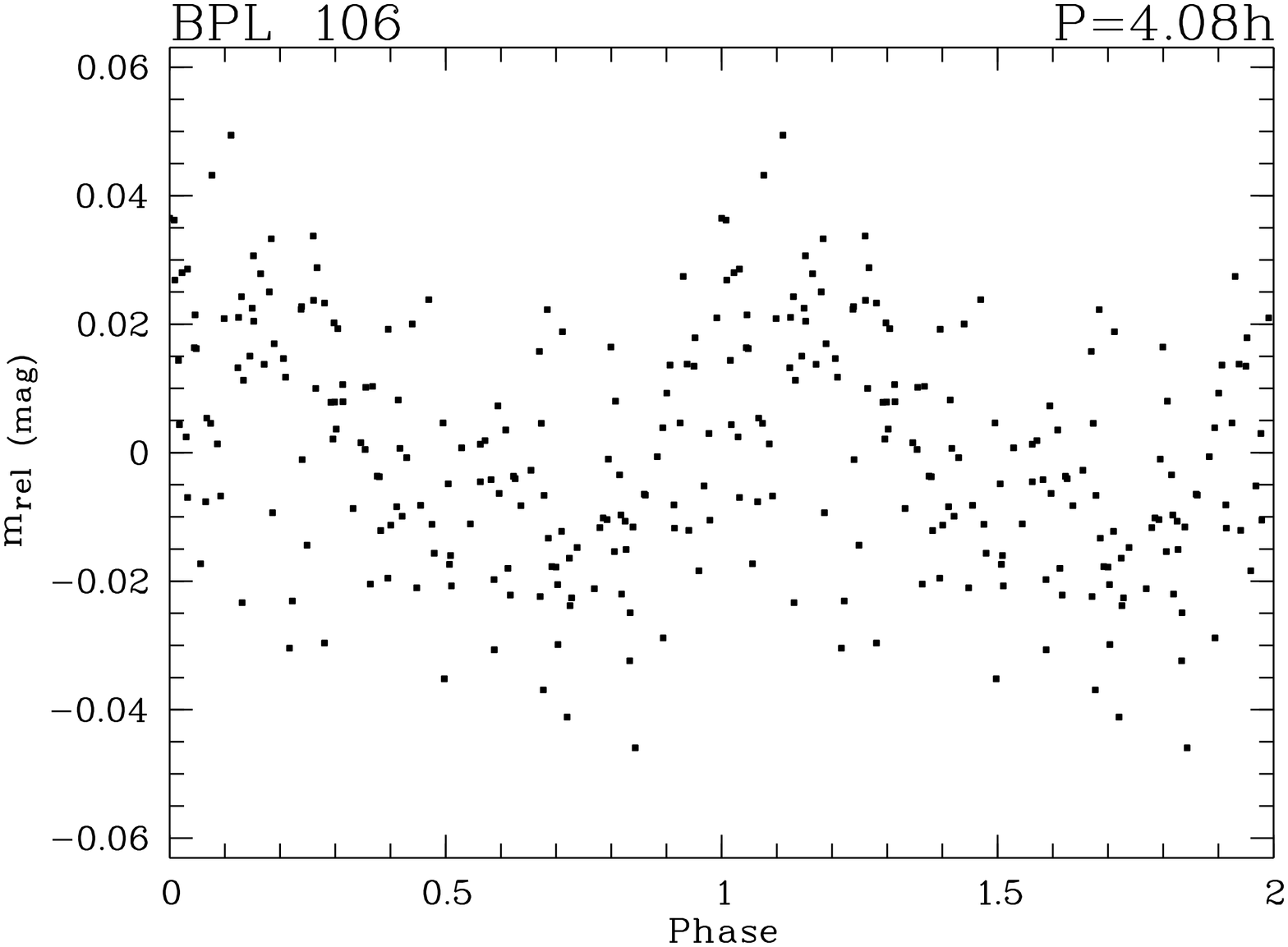}} \hfill
\resizebox{5.9cm}{!}{\includegraphics[angle=-90,width=6.5cm]{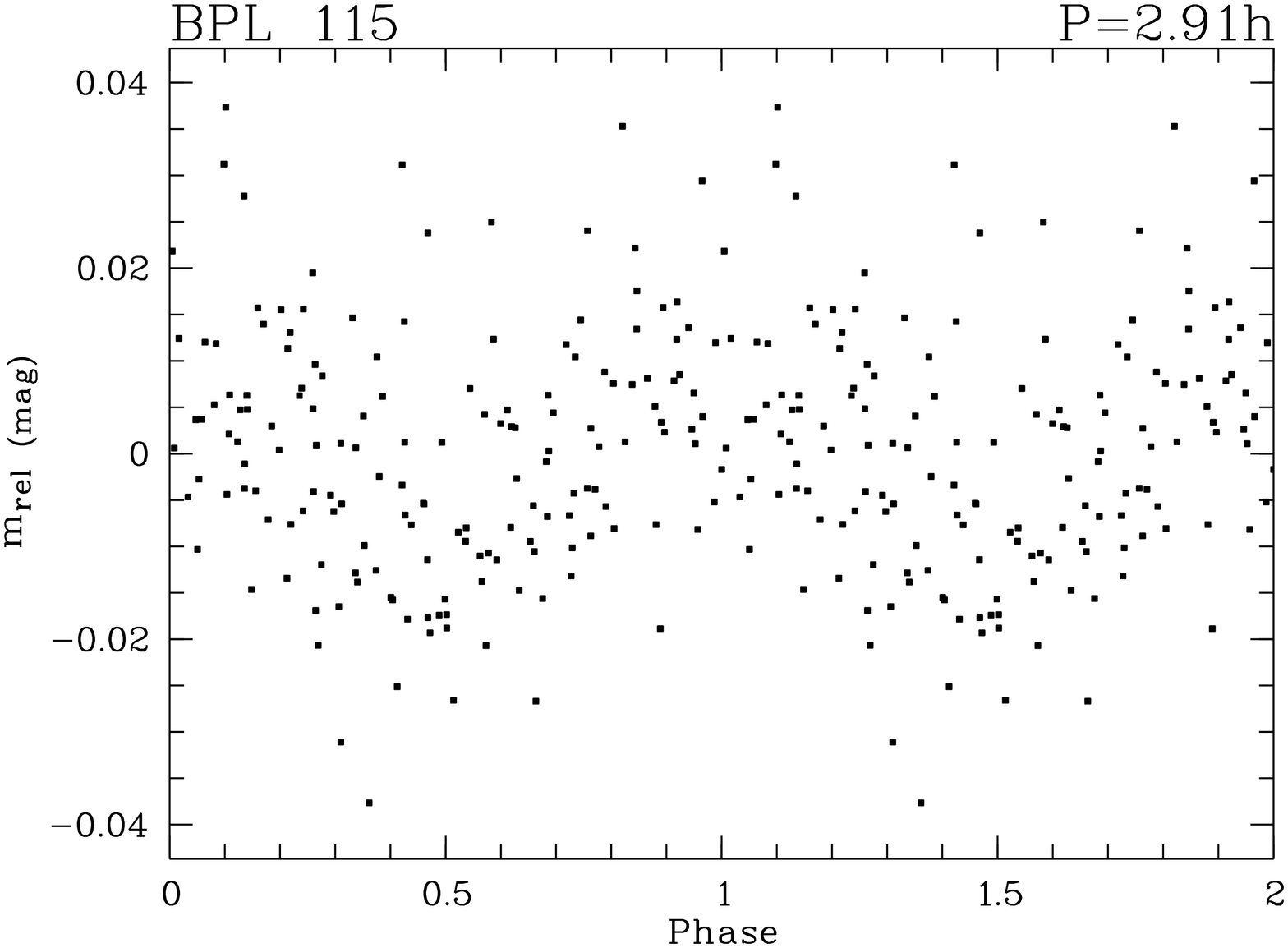}} \\
\resizebox{5.9cm}{!}{\includegraphics[angle=-90,width=6.5cm]{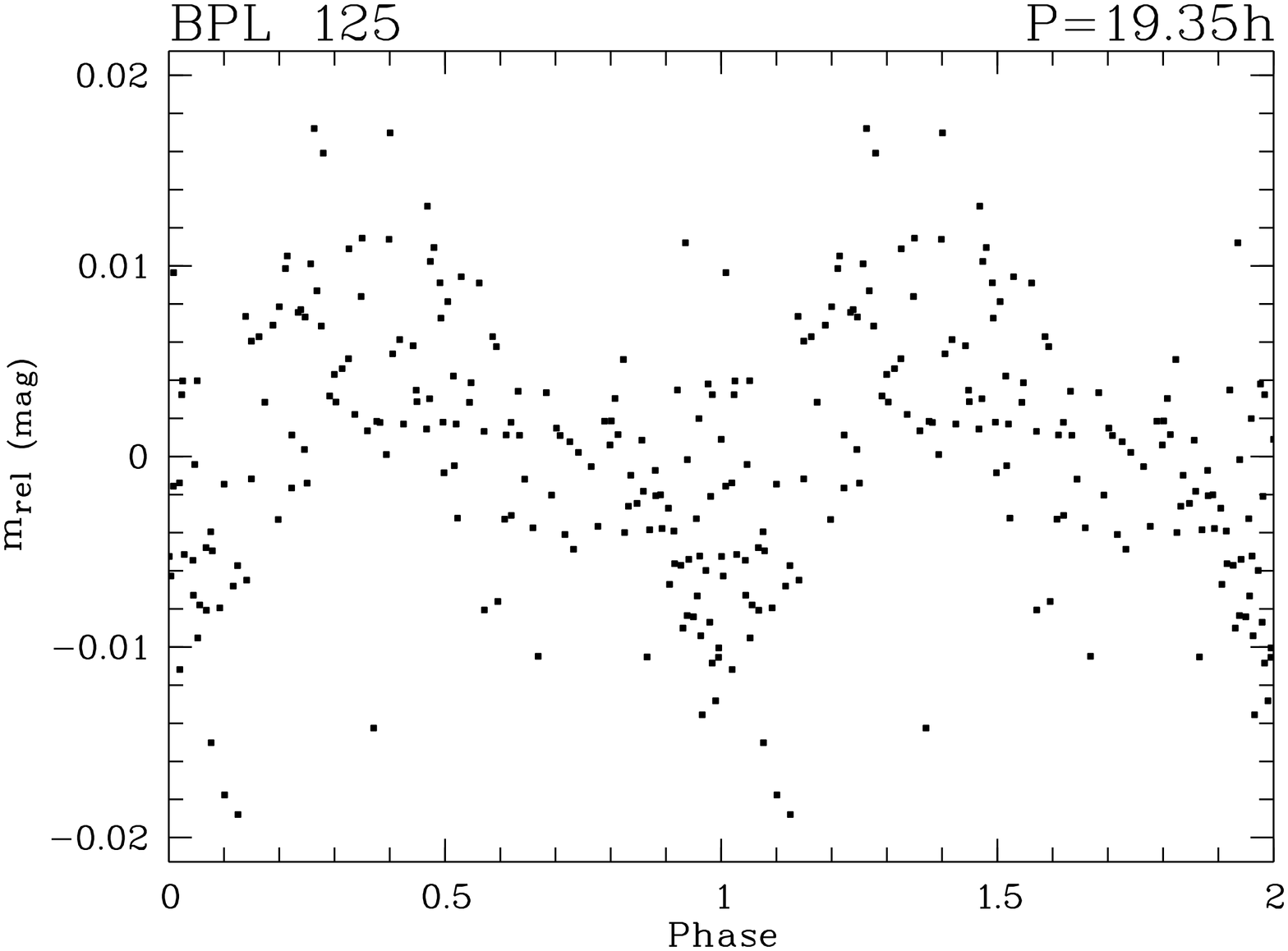}} \hfill
\resizebox{5.9cm}{!}{\includegraphics[angle=-90,width=6.5cm]{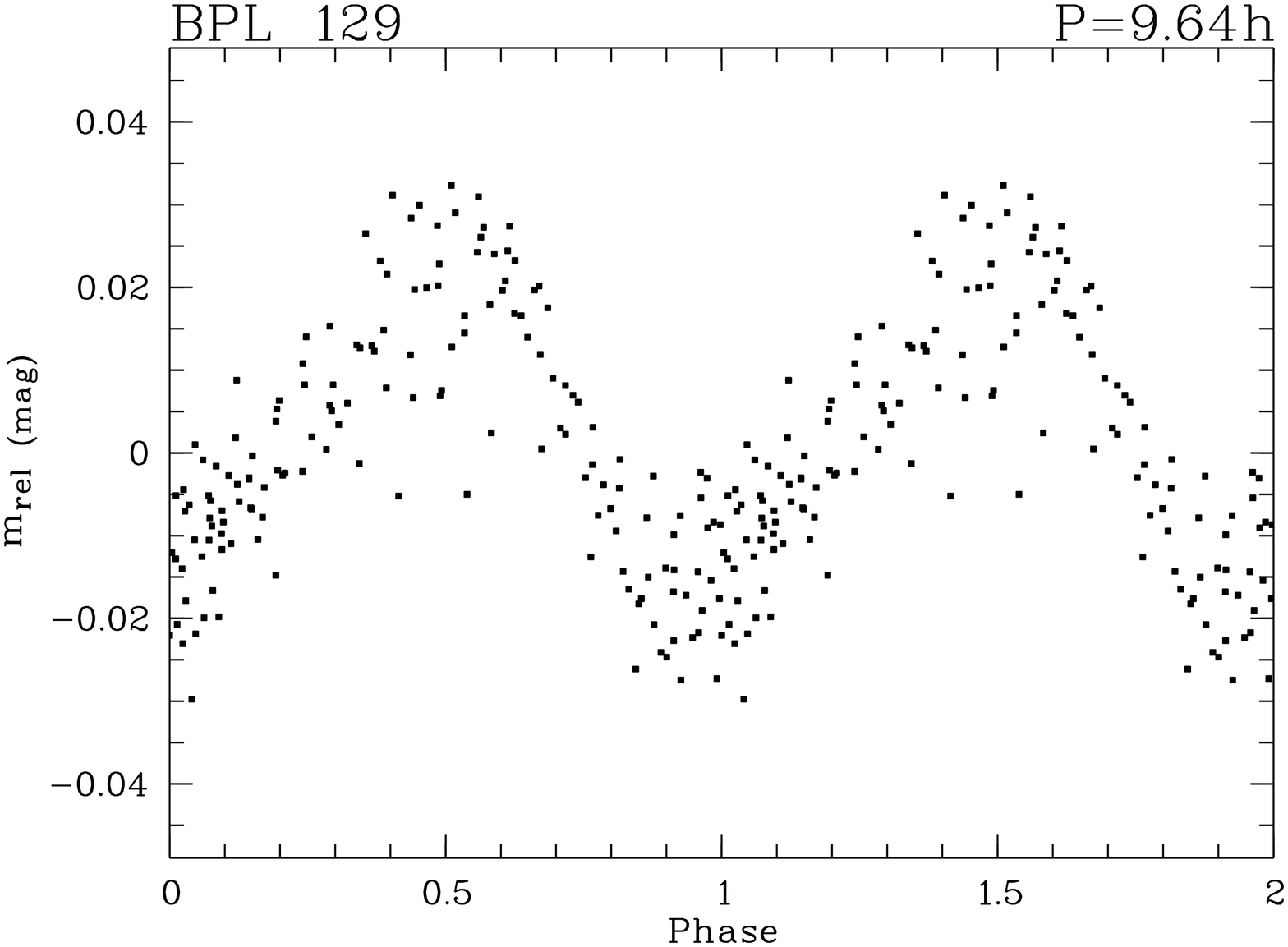}} \hfill
\resizebox{5.9cm}{!}{\includegraphics[angle=-90,width=6.5cm]{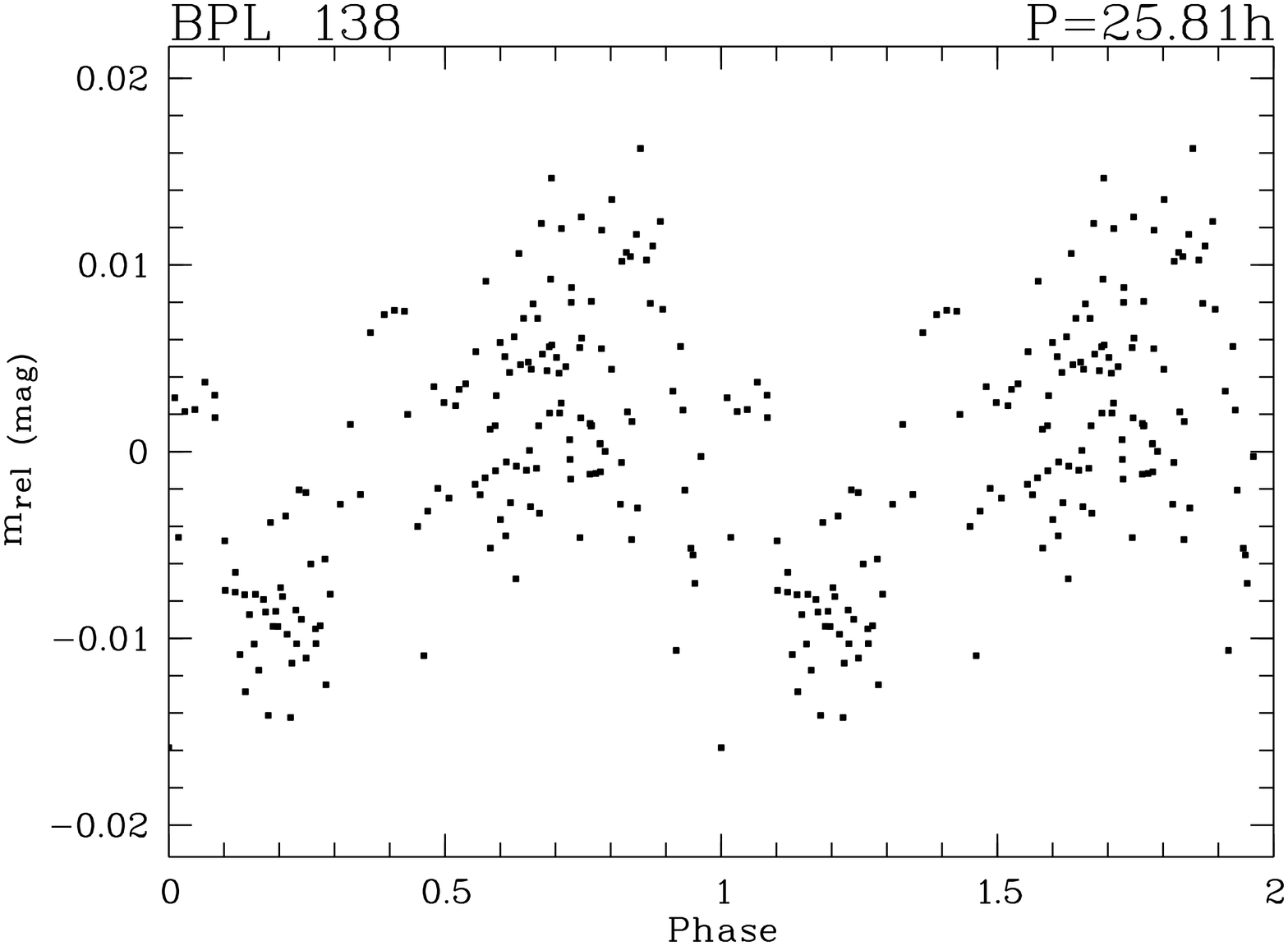}} \\
\resizebox{5.9cm}{!}{\includegraphics[angle=-90,width=6.5cm]{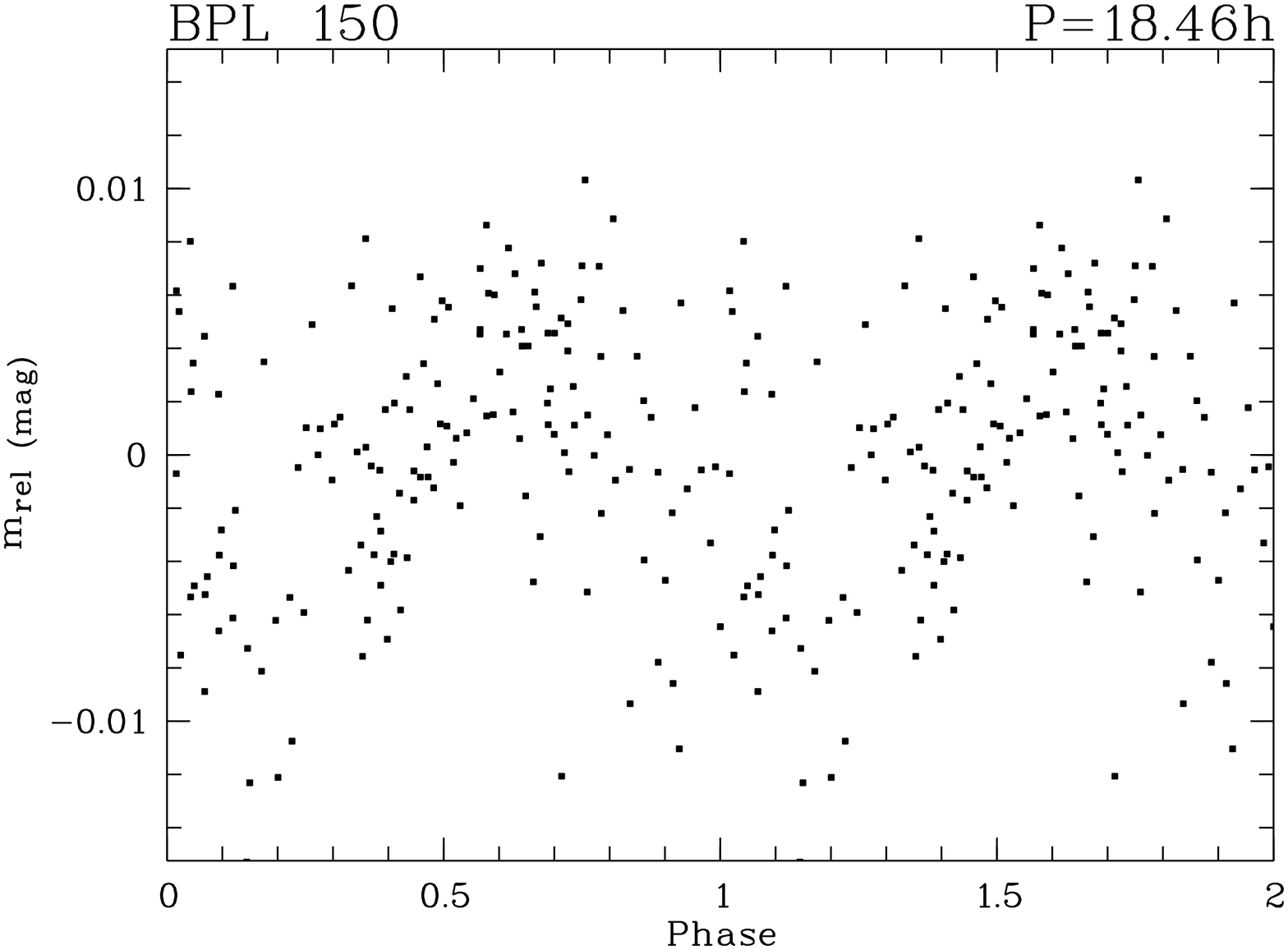}} \hfill
\resizebox{5.9cm}{!}{\includegraphics[angle=-90,width=6.5cm]{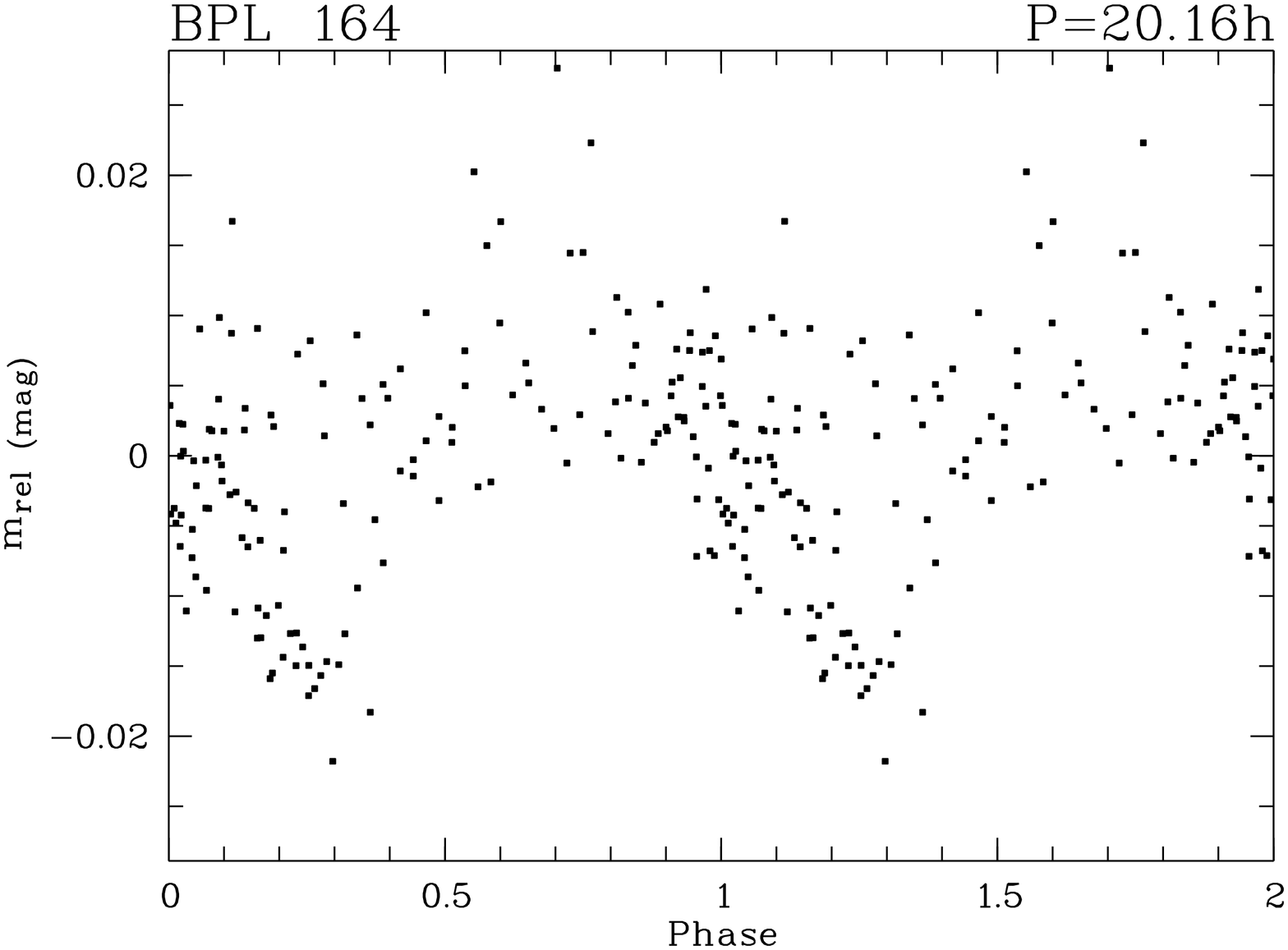}} \hfill
\resizebox{5.9cm}{!}{\includegraphics[angle=-90,width=6.5cm]{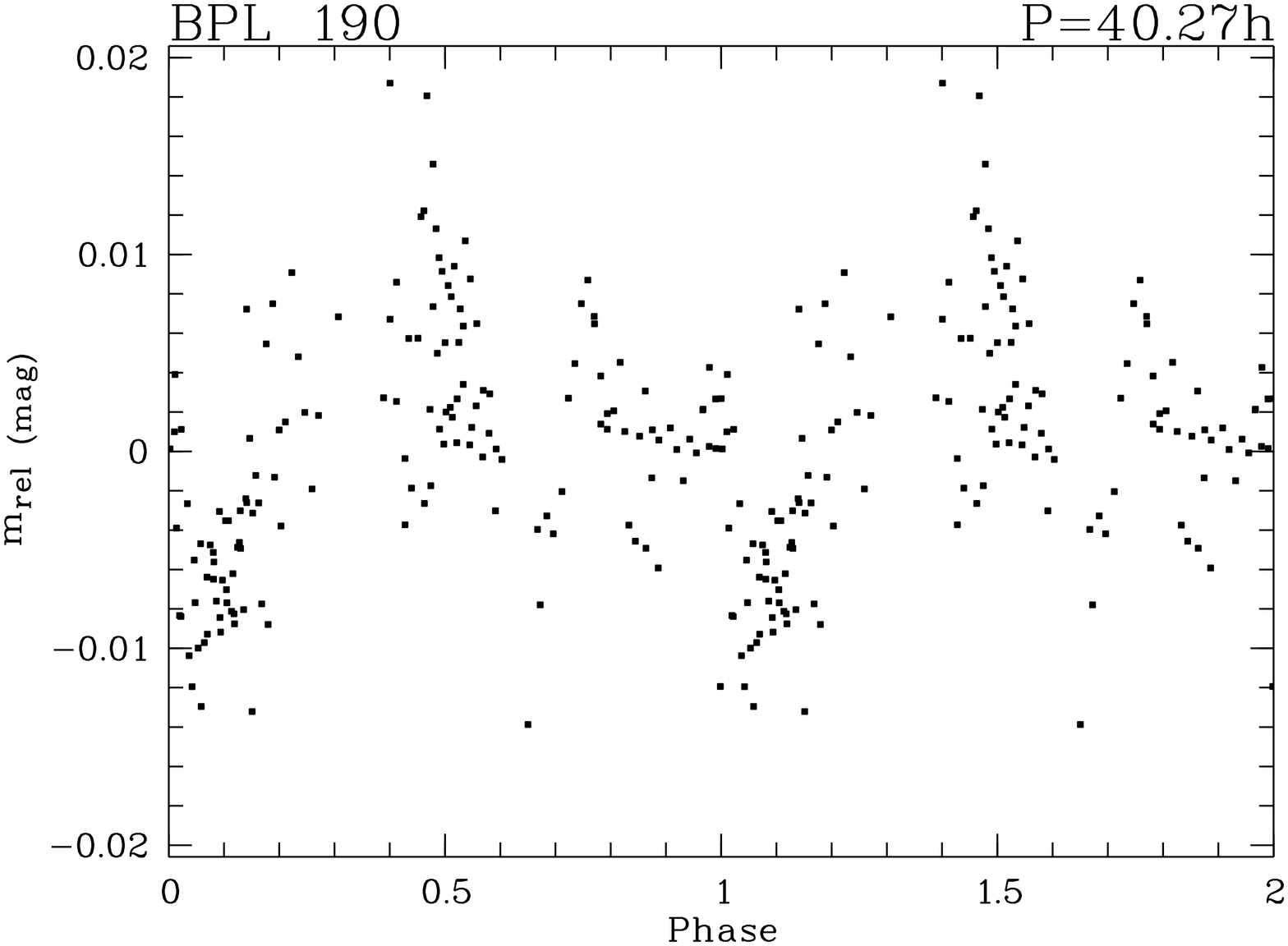}} \\
\caption{Phased lightcurves for the detected periodicities. No. and period from 
Table\,\ref{periods} are indicated.}
\label{phase}
\end{figure*}

a) The Scargle periodogram shows a peak exceeding the 99\% significance level,
as given by Horne \& Baliunas (\cite{hb86}). This gives us a first 
estimate for the False Alarm Probability (FAP) of the period.

b) The period is confirmed by the CLEAN algorithm, i.e. the highest peak
in the CLEANed periodogram corresponds to the detected period. This assures that
the period is not due to sidelobes and aliases, caused by the window function of
the data.

c) The scatter in the lightcurve is significantly reduced after the subtraction
of the (sine-wave approximated) period. (For the comparison of the scatter
we used the statistical F-test.)

d) Periodogram and phased lightcurve of at least ten nearby reference stars do not 
show the same period as the candidate. This makes sure that the detected period
is an intrinsic property of the candidate lightcurve.

e) The periodicity is visible in the phased lightcurve. We note that this
last test is the only subjective test among our period search criteria. It has
to be treated with caution, because our time series are composed of 150 or more 
data points. This large number of data points enables a reliable period detection, 
even if the noise level is very high (see Sect. \ref{sensitivity}).

\begin{table}[h]
\caption{Candidates with significant periodic variability (see text for
explanations).}
\begin{tabular}{ccccccc} \hline
BPL & M($M_{\odot}$) & P(h) & $\Delta$P(h) & A(mag) & FAP$\mathrm{_{E}}$(\%) & N\\
\hline
 102 & 0.25 & 21.4  & 0.49 & 0.013 & $<0.01$ & 158 \\
 106 & 0.08 & 4.08  & 0.02 & 0.032 & $<0.01$ & 159 \\
 115 & 0.10 & 2.91  & 0.01 & 0.014 & $0.01$ & 160 \\
 125 & 0.15 & 19.35 & 0.34 & 0.012 & $<0.01$ & 159 \\
 129 & 0.13 & 9.64  & 0.10 & 0.034 & $<0.01$ & 160 \\
 138 & 0.25 & 25.81 & 0.58 & 0.015 & $<0.01$ & 159 \\
 150 & 0.18 & 18.46 & 0.32 & 0.008 & $0.02$ & 152 \\
 164 & 0.13 & 20.16 & 0.53 & 0.021 & $<0.01$ & 150 \\
 190 & 0.15 & 40.27 & 1.57 & 0.012 & $<0.01$ & 153 \\
\hline				     	      
\end{tabular}			     
\label{periods}			     
\end{table}		

The final FAPs for the periods were determined using the bootstrap approach 
following K{\"u}rster et al. (\cite{ksc97}). For each candidate, we generated 
10000 randomized lightcurves by retaining the observing times and randomly 
redistributing the observed relative magnitudes amongst the observing times. The Scargle
periodogram was calculated for each of these randomized datasets, and the power
of the highest peak was recorded in each of them. The FAP is the fraction of datasets 
for which the power of the highest peak exceeds the power of the periodicity in the 
observed lightcurve. It turned out that these empirical values for the FAP are very 
similar to our first FAP estimate from the peak height in the Scargle periodogram. We
accepted a period if the FAP is below 1\%. For a more detailed discussion of
the validity of the bootstrap approach, we refer to SE2004.

We established periodic variability for nine targets. Eight of these objects are
also variable according to the generic variability test in Sect. \ref{gen}).
The period search is more sensitive than the simple variability test
of Sect. \ref{gen}, therefore it detects one more low amplitude variable (object 
BPL150). The periods range from 3 to 40 hours, the amplitudes from 0.01 to 0.03\,mag. 
In Table \ref{periods}, we list all relevant data for these objects. The 
first column gives the object no. according to the nomenclature of Pinfield et 
al. (\cite{phj00}). Our periods were determined by fitting the CLEANed periodogram peak 
with a Gaussian. Period errors ($\Delta$P) are based on the half width at half 
maximum of the fitted Gaussian, transformed to time space. The amplitudes (A) 
correspond to the peak-to-peak-range of the binned lightcurve. N is the number 
of data points used for the period search. The masses in column 2 are estimated 
by transforming the $I_{KP}$-band magnitudes of Pinfield et al. (\cite{phj00}) in 
the $I_C$ band, using the transformation given in Pinfield et al. (\cite{phj00}).
These magnitudes were then compared with the Pleiades evolutionary tracks of Baraffe 
et al. (\cite{bca98}), adopting $m-M=5.5$ and an age of 125\,Myr.
The phased lightcurves of all objects with detected period are shown in Fig. 
\ref{phase}.

For several objects with significant periodic variability, the phased 
lightcurve of Fig. \ref{phase} shows a relatively high noise level. As 
noted above, this does not necessarily mean that the period 
detection is not reliable, because of our large number of data points. However, 
it could be an indication for the evolution of surface properties. If the surface 
pattern is constant, the amplitude of the periodic signal will be constant. On the 
other hand, if the surface pattern evolves, the amplitude of the periodicity may
vary.

\begin{figure}[htbd]
\centering
\resizebox{\hsize}{!}{\includegraphics[angle=-90,width=6.5cm]{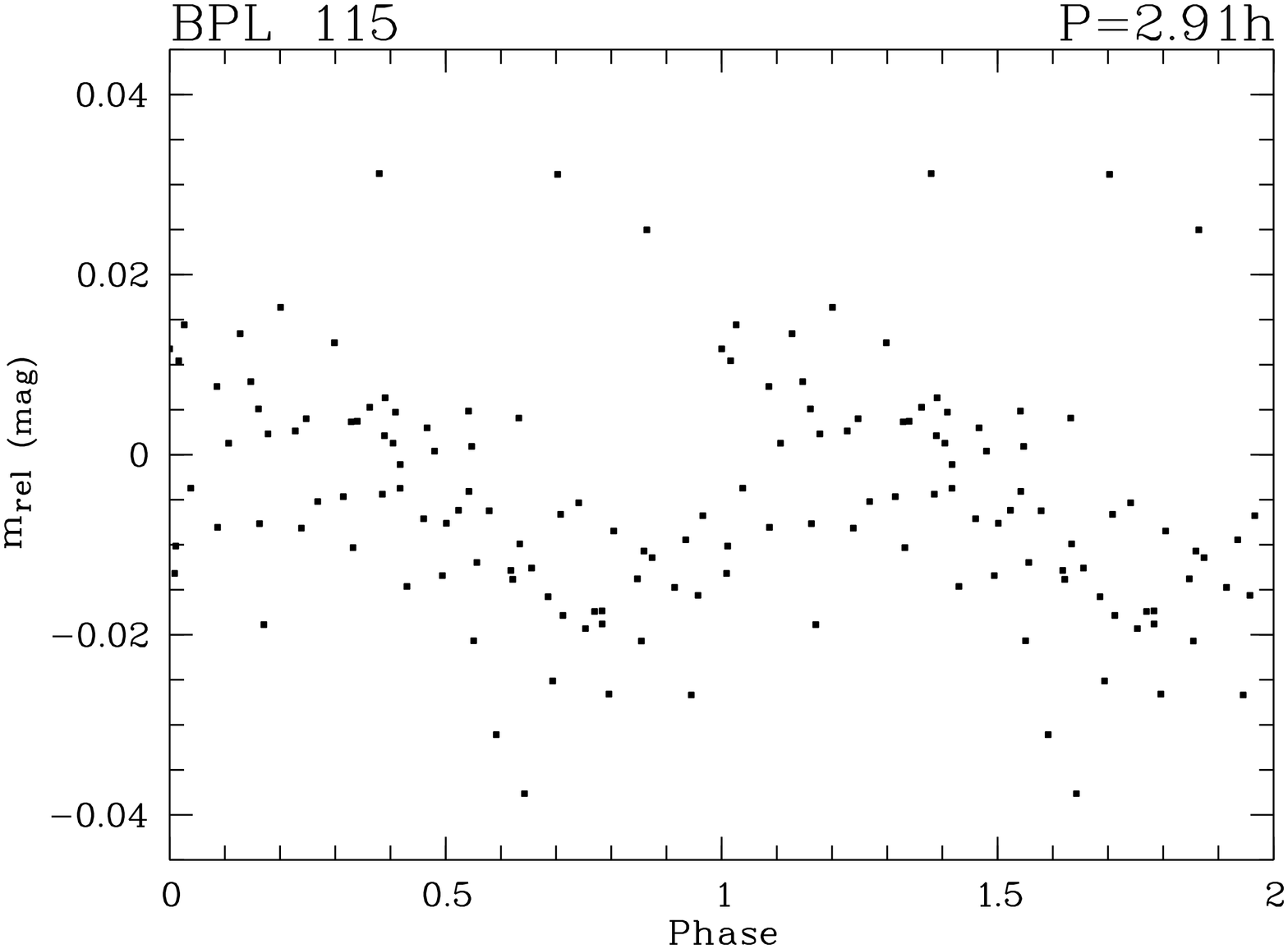}}
\resizebox{\hsize}{!}{\includegraphics[angle=-90,width=6.5cm]{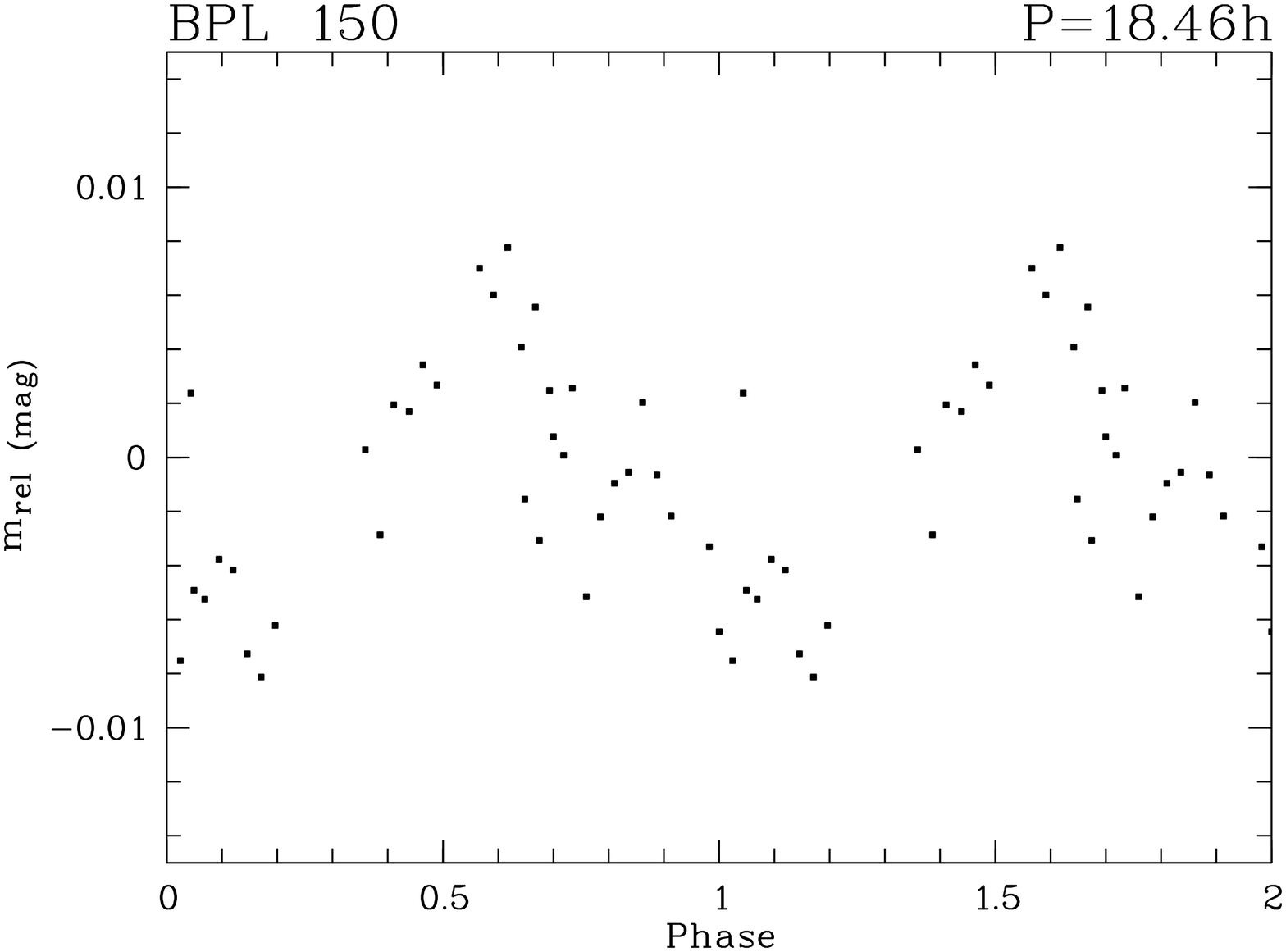}}
\caption{Phased lightcurves for part of the time series of objects 115 (upper panel,
only datapoints after 10 Oct UT12:00) and 150 (lower panel, only datapoints before
6 Oct UT12:00)}. 
\label{lc300}
\end{figure}

To investigate the surface pattern stability, we examined the phased
lightcurves of the periodic objects for parts of the time series.
Five out of nine objects, namely the targets 106, 115, 164, 150, 
190, show clear evidence for spot evolution in the course of our 18-night
observing run. For these objects, the period is obvious in the first or last 
part of the time series, but relatively noisy over the whole dataset. In Fig. 
\ref{lc300}, we show exemplarily the phased lightcurves for objects 115 and 150 
for a part of the time series. In both cases, the period is much more evident than in 
Fig. \ref{phase}.

\subsection{Sensitivity}
\label{sensitivity}

\begin{figure}[htbd]
\centering
\resizebox{\hsize}{!}{\includegraphics[angle=-90,width=6.5cm]{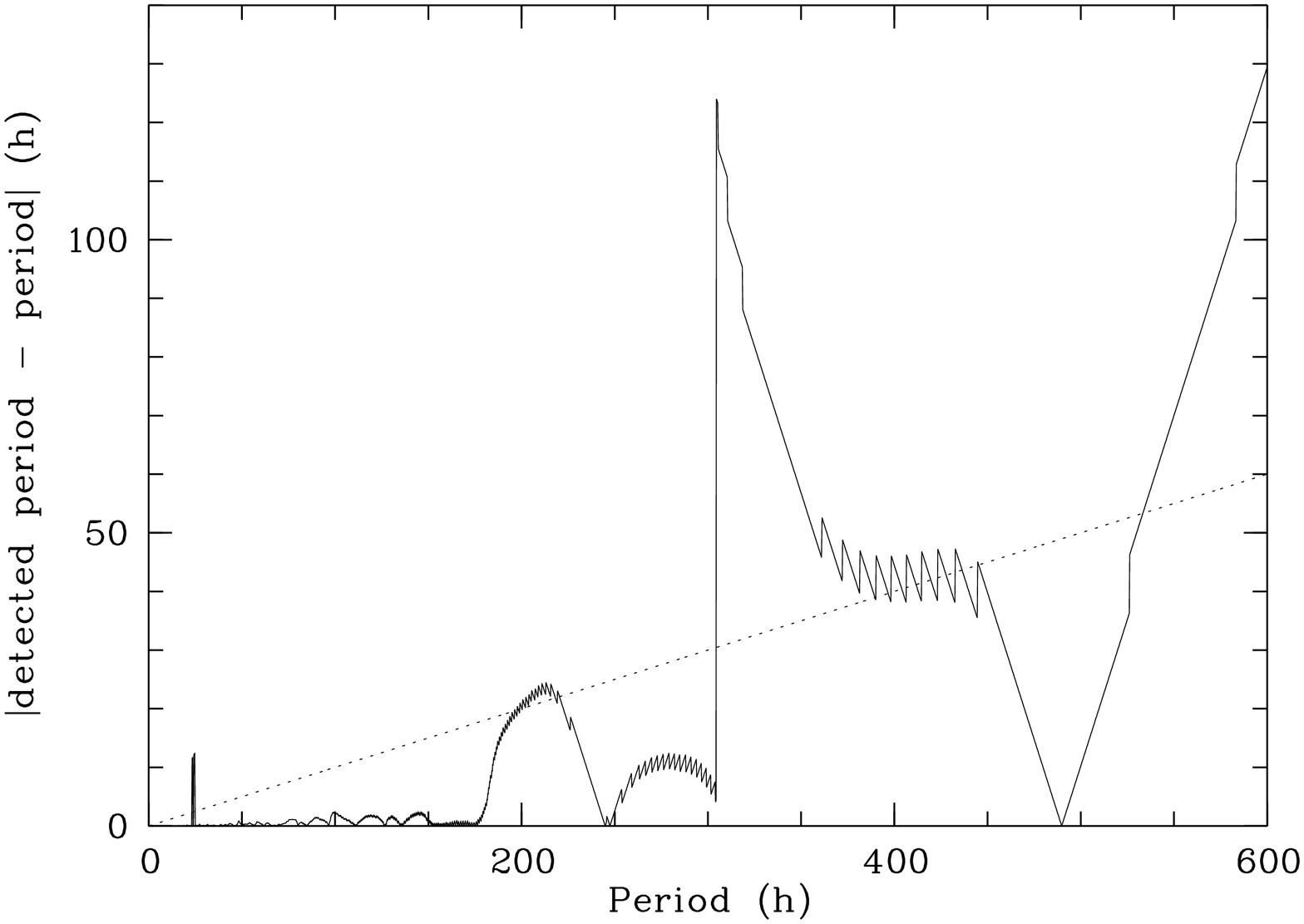}}
\resizebox{\hsize}{!}{\includegraphics[angle=-90,width=6.5cm]{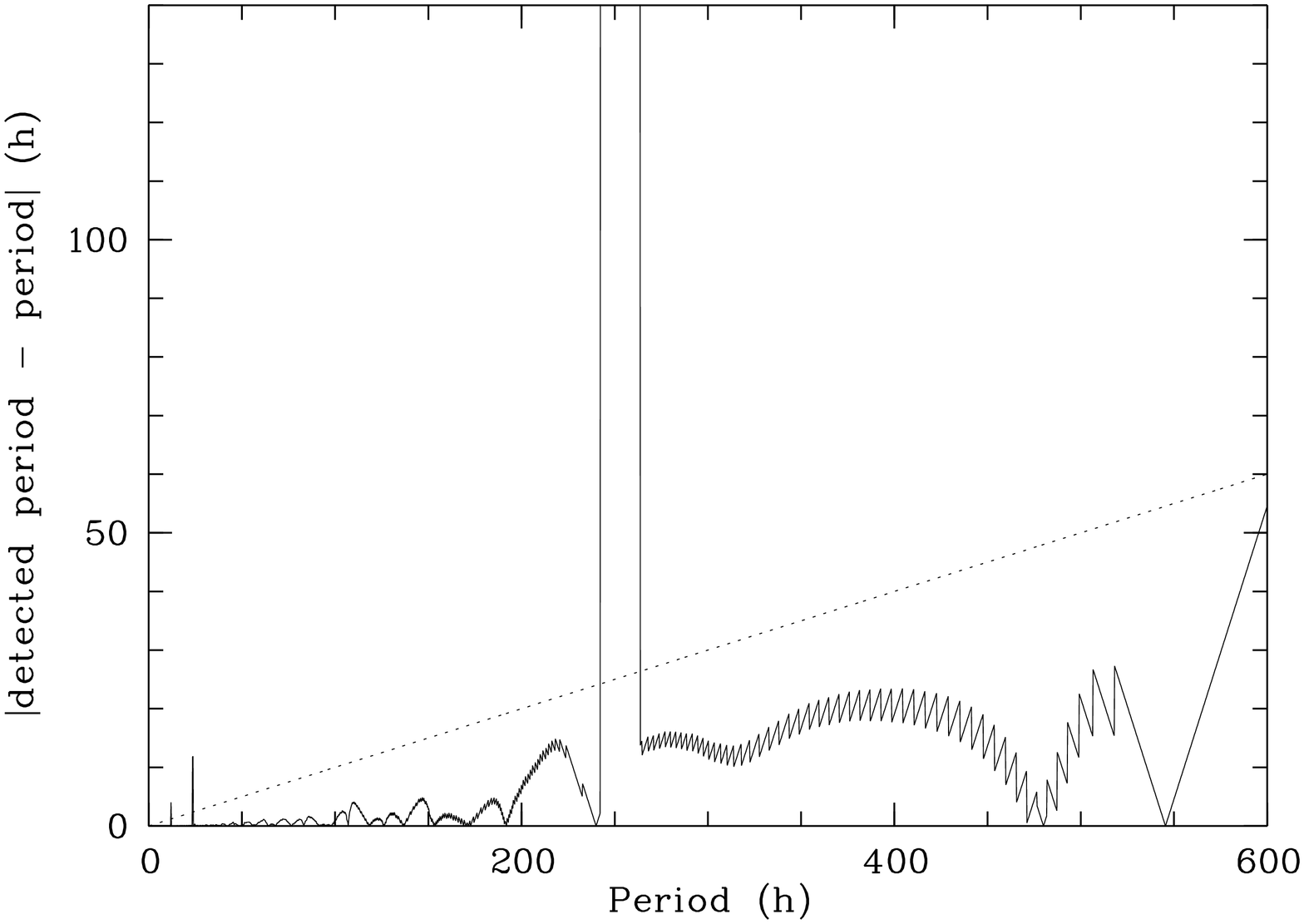}}
\caption{Sensitivity of the period search: The absolute difference between detected period
  and true period vs. true period for field A (upper panel) and field B (lower panel). The 
  signal-to-noise ratio of the periodicity is 2.5; the dotted line corresponds to a period 
  error of 10\%.}
\label{sens}
\end{figure}

From our time sampling (see Fig. \ref{sampling}), we estimate that we are sensitive 
to periods from 0.5\,hours up to 18 days. To assess the sensitivity of our period
search more precisely, we executed simulations similar to those described in SE2004. We 
selected non-variable objects from our data and added sine-shaped periodicities to their
lightcurves, so that the signal-to-noise ratio (defined as ratio of period amplitude
and scatter in the original lightcurve) is similar to the periodicities in
Table \ref{periods}. We computed the Scargle periodogram for periods between
0.01\,h and 300\,h and recorded the frequency of the highest peak. The absolute 
difference between the imposed period and the detected period serves as an indicator
for the reliability of our period search. Since the sampling is slightly different for both 
fields, we did the simulation for both fields separately. The results are shown in Fig. 
\ref{sens}. The reliability of the period search is slightly variable, but the uncertainty is
below 10\% up to periods of 300\,h, with two exceptions: a) In both diagrams, there is 
a small peak at one day, caused by the regular gaps between the observing nights, and
b) for field A, there is a narrow window of non-sensitivity between periods of 240
and 270\,h. This can be explained with the lack of data points between days 6 and 11 in
Fig. \ref{sampling} (lower panel). 

From this simulation, we can be confident that we are sensitive to photometric periods 
between 0.01\,h and 300\,h, with no major biases. For longer periods, the uncertainty 
of the period determination generally increases, and there is a large gap of non-sensitivity
for field A around periods of 320\,h, probably caused by the lack of data points between
days 13 and 17 in Fig. \ref{sampling} (upper panel). Therefore, the period search
might be not sensitive for $P>300$\,h. Since we do not observe any period between 50\,h
and 235\,h, a range where our period search is reliable within $\pm 10$\%, it is,
however, very unlikely that there exist many periods $>300$\,h in our lightcurves.

Furthermore, we investigated the sensitivity of our period search as a function of
signal-to-noise ratio. This time, we fixed the period to $P=18$\,h, a value typical for
our detected periodicities, and varied the amplitude of the added sine wave. Again,
we computed the Scargle periodogram and recorded the frequency of the highest peak.
The difference between imposed period and detected period is larger than 10\,h
for very small amplitudes, but decreases to values $<0.04$\,h for signal-to-noise $>1.05$. 
Thus, our period search is sensitive down to signal-to-noise ratios of about 1.05. On the 
other hand, the minimum signal-to-noise ratio of our periods in Table \ref{periods} is 1.27 
(object BPL115), making us confident that our periods are reliable.

\section{Origin of the observed variability}
\label{ori}

Following the usual interpretation, we attribute the observed periodic
variability in the lightcurves to the existence of surface features 
co-rotating with the objects, which must be asymmetrically distributed 
to induce a photometric variation. The surface features could arise from 
two fundamentally different processes, dust condensation and magnetic 
activity, which will be discussed in the following.

Recently, several groups reported about photometric variability of 'ultracool 
dwarfs' in the field, i.e. VLM objects with spectral types L or T
(Tinney \& Tolley \cite{tt99}, Bailer-Jones \& Mundt \cite{bm01}, Mart\'{\i}n et 
al. \cite{mzl01}, Clarke et al. \cite{cot02}, \cite{ctc02}, Gelino et al. 
\cite{gmh02}, Enoch et al. \cite{ebb03}). These objects are too cool to 
generate magnetically induced spots (Gelino et al. (\cite{gmh02}), but 
cool enough to form dust clouds in the atmosphere (Allard et al. \cite{aha01}), 
which most probably are the origin of the observed variability.

Young VLM objects have significantly higher effective temperatures
than ultracool dwarfs. E.g., our Pleiades VLM stars have $T_{eff}>2800$\,K 
(Baraffe et al. \cite{bca98}), corresponding to spectral types earlier
than M7. For these M-type objects, the existence of dust clouds is 
unlikely, because their spectra and near-infrared colours are well-approximated
by dust-free models (Delfosse et al. \cite{dfs00}, Dawson \& DeRobertis 
\cite{dd00}). On the other hand, they are capable of sustaining magnetic 
activity (Delfosse et al. \cite{dfp98}, Mohanty \& Basri \cite{mb03}).  
This lead us to the conclusion that the photometric variability on
our targets is caused by magnetically induced spots.

Attributing the photometric variability to magnetic spots, the properties 
of the variability give us important constraints on the spot properties:

\noindent
a) From our data, we see evidence for temporal evolution of the spot 
patterns. On the one hand, we find two objects (BPL111 and BPL128) with 
significant variability, but without period detection, whose lightcurve 
variation is most likely diluted by surface evolution (see Sect. \ref{gen}). 
On the other hand, five out of nine periodic objects show clear indications for 
spot evolution (see Sect. \ref{per}). Altogether, there are eleven targets
with variability probably caused by spot activity, from which seven
show evolving surface patterns. The 'half-life period' of the spot pattern, 
i.e. the time in which half of the objects show evolved surface features, is 
thus $\tau_{1/2} = 0.5 \times 11/7 \times 18$\,d $= 14$\,d. This is in clear 
contrast to the results for the cloudy L dwarfs, where the surface features seem 
to be variable on timescales of hours, and thus prevent the detection of 
significant periods in the lightcurve (e.g., Bailer Jones \& Mundt \cite{bm01}).

\noindent
b) We compared the photometric amplitudes of the variations with
the results of studies of more massive Pleiades members. As comparison sample, 
we used the periods from the Open Cluster Database (see Sect. \ref{massper}
for a description of this sample). We note that the amplitudes in the literature 
are mostly defined as difference between maximum and minimum in the original
time series, whereas we measured the amplitude in the binned phased lightcurve.
However, since the lightcurves of solar mass stars in the literature
show mostly very low scatter, binning should decrease their amplitudes only 
marginally. On the other hand, the VLM lightcurves are highly scattered
(see Fig. \ref{phase}). Therefore, the peak-to-peak amplitudes measured from
the raw lightcurve are essentially determined by the photometric noise, and
thus overestimate the real amplitude. Binning reduces the noise and delivers 
amplitudes, which can be compared with literature values.

\begin{figure}[h]
\centering
\resizebox{\hsize}{!}{\includegraphics[angle=-90,width=6.5cm]{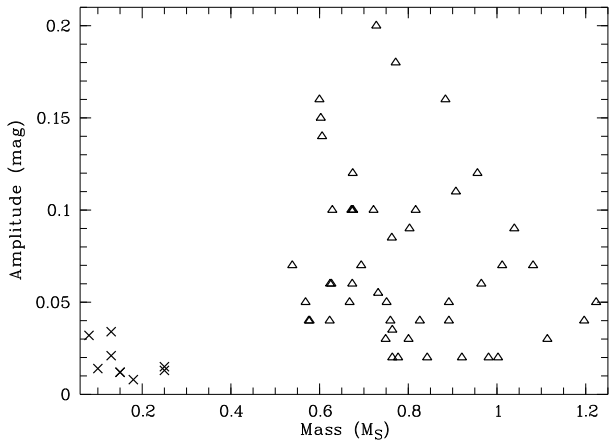}}
\caption{Photometric amplitude vs. mass for stars from the Open Cluster
Database (triangles) and our targets (crosses). The detection limit for
the solar-mass stars is 0.02\,mag, explaining the lack of stars with
very low amplitude in this sample.}
\label{ampmass}
\end{figure}

From Fig. \ref{ampmass} it is obvious that solar mass stars show more frequently 
photometric variations with $A>0.05$\,mag. According to a $\chi^2$ test, the 
amplitude distributions of both samples are significantly different ($\mathrm{FAP}<0.01$\%). 
The lack of amplitudes larger than 5\% on VLM objects can be explained with either a 
smaller relative spotted area, a more symmetric spot distribution, or lower contrast 
between spots and photospheric environment. The latter point seems to be more 
probable, since simulations show that the intensity contrast of the granulation 
pattern on M dwarfs is decreased by a factor of 14 in comparison with the Sun 
(Ludwig et al. \cite{lah02}). Symmetric spot distribution {\it and} low contrast, 
however, are predicted properties of a stellar surface governed by a turbulent 
dynamo, which might be the origin of magnetic activity on fully convective VLM 
objects (Durney et al \cite{ddr93}).

\noindent
c) Compared with more massive stars, the fraction of objects with measurable
photometric variation is significantly reduced among VLM objects.
From our 26 targets, only 9 (35\%) show significant periodic variability. In 
comparison, Krishnamurthi et al. (\cite{ktp98}) found periods for 21 out of
36 (i.e. 58\%) stars in the Pleiades with masses from 1.2 to 0.5\,$M_{\odot}$. 
Although these values might be biased by the incompleteness of the period search 
or the target selection (e.g., cluster members selected by X-ray activity), they 
support the notion for low spot coverage, low temperature contrast and/or symmetric 
distributions of spots on VLM objects, as outlined above.

\section{The mass-period relation}
\label{massper}

To investigate the mass dependence of the rotation, we first compare our derived
rotation periods with those of similar studies for more massive stars. From the Open 
Cluster Database, we collected a sample of rotation periods for solar mass stars in 
the Pleiades (see Sect. \ref{intro} for complete references). These stars have spectral 
types G, K, or early M, corresponding to masses from 1.2 to 0.5\,$M_{\odot}$, and are 
thus complementary to our targets. We estimated masses for these comparison stars by 
comparing the V-band magnitudes with the 125\,Myr isochrone of Baraffe et al. 
(\cite{bca98}), the same that we used to estimate masses for our VLM objects 
(Sect. \ref{per}). Therefore, we are confident that the absolute masses can be
compared with each other, even if they are systematically offset because of an
over- or underestimate of the assumed age of the Pleiades or shortcomings of 
the models.

\begin{figure}[htbd]
\centering
\resizebox{\hsize}{!}{\includegraphics[angle=-90,width=6.5cm]{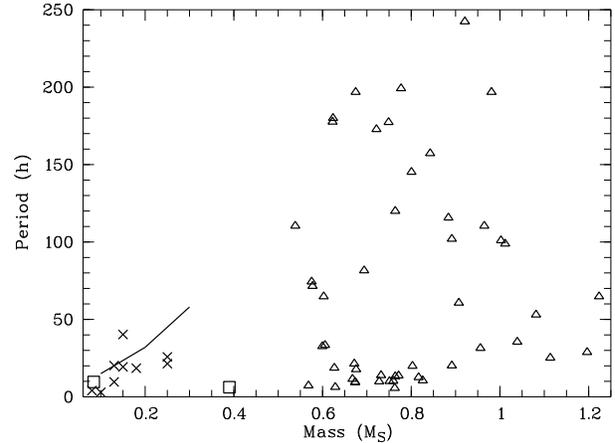}}
\caption{Rotation periods vs. mass: Our VLM rotation periods are shown as crosses. 
Triangles mark the periods for more massive stars from the Open Cluster Database
(see Sect. \ref{intro} for complete references). The two squares show the periods 
from Terndrup et al. (\cite{tkp99}). The solid line marks a rough lower envelope
to the observed $v \sin i$ values of Terndrup et al. (\cite{tsp00}).}
\label{allper}
\end{figure}

In Fig. \ref{allper}, we plot period vs. mass for our VLM objects (crosses), the 
solar-mass stars in the literature (triangles), and the two VLM stars from Terndrup 
et al. (\cite{tkp99}, squares). Whereas solar mass stars show rotation periods up to 
10 days, VLM objects have all periods below 2 days. However, we might have missed
slowly rotating objects, e.g. because of spot evolution or absence of spots on
these objects. For an independent evaluation of our upper period limit, we therefore
compared our results with those of Terndrup et al. (\cite{tsp00} and references 
herein) who collected a large sample of spectroscopic rotational velocities for 
low-mass members of the Pleiades. We extracted a rough lower envelope of 
$v \sin i=$\,7, 10, and 15\,kms$^{-1}$ for masses of 0.3, 0.2, 0.1$\,M_{\odot}$ 
from their Fig. 7. This lower $v\sin i$ envelope was transformed into an upper period 
envelope using the radii from the models of Chabrier \& Baraffe (\cite{cb97}). These 
upper period limits are shown in Fig. \ref{allper} as solid line. With one exception, 
all our periods lie below this line, and are thus in good agreement with the
$v \sin i$ data. Hence, all available data indicate a scarcity of slow rotators
among VLM objects.

We used the $\chi^2$ test to compare our periods with those measured for solar-mass stars. 
The null hypothesis 'VLM objects and solar mass stars show the same period distribution'
is rejected with a FAP of $<0.01$\%. Thus, the lack of slow rotators among 
VLM objects leads to a significant difference in the period distributions.

\begin{figure}[htbd]
\centering
\resizebox{\hsize}{!}{\includegraphics[angle=-90,width=6.5cm]{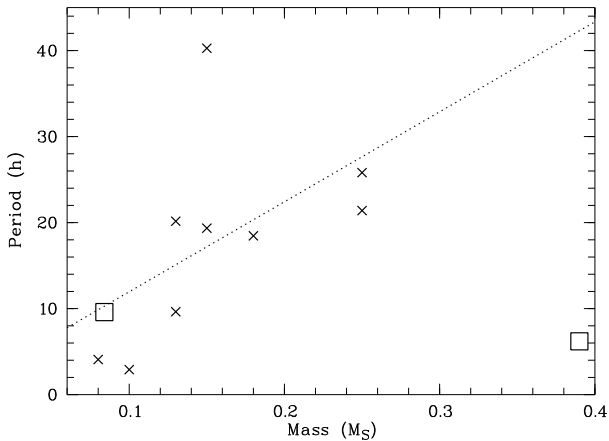}}
\caption{Period-mass relation in the VLM regime: Crosses are the nine periods 
determined in this work. The dashed line is a linear fit to their period-mass relation. 
The two squares indicate the two periods of Terndrup et al. (\cite{tkp99})}
\label{permass}
\end{figure}

A second aspect of the mass-period relationship can be seen in Fig. \ref{permass}, which is 
an enlargement of Fig. \ref{allper} for the VLM regime, containing period vs. mass for our 
nine VLM periods (crosses): Even in the VLM regime, the periods increase linearly with mass.
A linear least-square fit to this period-mass relation gives 
$P = (105 \pm 61)\,(M/M_{\odot}) - (1.5\pm 10)$\,h (dotted line in Fig. \ref{permass}). 
This correlation, however, is not yet very strong: The correlation coefficient is 0.54, leaving
a probability of 13\% that periods and masses are uncorrelated. It is thus clearly necessary 
to substantiate the tendency of faster rotation with lower masses in the VLM regime by more 
measurements. In Fig. \ref{permass}, we also show the two periods of Terndrup et al. 
(\cite{tkp99}, squares). One of these periods is in good agreement with the linear fit, 
while the second one (for the object HHJ-409) is a clear outlier. The period, however, is 
convincing, and the probability that this star is not a cluster member is quite low (Hambly 
et al. \cite{hhj93}). Certainly, its period needs reconfirmation.

The positive correlation between period and mass is confirmed by observations
in younger clusters. Herbst et al. (\cite{hbm01}) studied a large sample of
periods for stars down to 0.1$\,M_{\odot}$ in the ONC (age 1\,Myr). The median
of their periods in the mass range $0.1<M<0.35\,M_{\odot}$ decreases steadily 
with mass. In SE2004, we demonstrate for the very young $\sigma$\,Ori 
cluster (age 3\,Myr) that this relationship extends to substellar objects. The 
period median from these two studies can be fitted as 
$P = (491\pm 26)\,(M/M_{\odot}) - (7.9\pm 5.9)$\,h, 
i.e. the slope is much steeper than for the Pleiades VLM stars. The flatter 
relationship for the Pleiades can be understood as a consequence of the 
pre-main sequence contraction process: In the course of their evolution from 
the age of $\sigma$\,Ori to the age of the Pleiades, the radii of the VLM objects 
decrease, and therefore their rotation accelerates (see Sect. \ref{model} for
a detailed discussion of this process). The relative decrease of
the rotation period is, however, only little dependent of mass. Hence, 
the period-mass relationship becomes flatter.

We note that for the ONC and the $\sigma$\,Ori samples, the 
correlation is only apparent from the {\it median} of the periods, whereas in 
the case of the Pleiades the trend can be seen directly from the periods themselves. 
Since the positive correlation between rotation and mass is already present at 
very young ages, it must be produced in the earliest phases of rotational evolution, 
e.g. by mass-dependent angular momentum loss.

\section{Rotational evolution of VLM stars}
\label{ageper}

\subsection{Modelling the rotational evolution of VLM stars}
\label{model}

The periods for VLM objects in the Pleiades (from this work) 
and in the $\sigma$\,Ori cluster (from SE2004) will now be combined to deliver 
constraints for models of angular momentum evolution in the VLM regime. 
Our goal is to see if a simple model can reproduce the period distribution in the 
Pleiades by transforming the $\sigma$\,Ori periods to the Pleiades age. This 
transformation should take into account the basic ingredients of angular momentum 
evolution. We use the following nomenclature: The initial period of a given object 
in the $\sigma$\,Ori cluster will be called $P_i$, and the corresponding evolved period 
for the Pleiades age $P_e$. Similarly, we will call the radii in the $\sigma$\,Ori
cluster and in the Pleiades $R_i$ and $R_e$. For the ages of the $\sigma$\,Ori 
cluster and the Pleiades, we use $t_i$ and $t_e$. Note that our period search 
in the Pleiades is highly sensitive up to $P=300$\,h, as shown in Sect. \ref{sensitivity}, 
whereas the study in $\sigma$\,Ori is hampered by its less favourable time 
coverage, causing several windows of decreased sensitivity for $P>30$\,h. Therefore, 
the transformation is done {\it forward} in time, to avoid unnecessary biases.

The rotational evolution of low-mass stars is, according to the current paradigm,
determined by four factors:

\noindent
a) The evolution starts with a given initial angular momentum. We use the period 
distribution in $\sigma$\,Ori as starting point for the model.

\noindent
b) Star-disk interactions probably play an important role for angular momentum
regulation in the T Tauri phase. According to the so-called disk-locking paradigm
(Camenzind \cite{c90}, K{\"o}nigl \cite{k91}),
young stars are locked to their disks and thus cannot spin up as long as the disk
is not dissipated. However, because of inconsistent observational results, the 
disk-locking scenario remains controversial (see Stassun \& Terndrup \cite{st03} 
for a detailed discussion). In SE2004, we showed that accretion disks are rare 
in the $\sigma$\,Ori cluster. Therefore, for our purposes disk-locking to first 
order is neglectable, but we will keep in mind that it might play a role for some 
objects.

\noindent
c) Rotation rates will be influenced by the contraction and the changes of the 
internal structure of the star. The latter point, however, can be neglected for 
fully convective objects, since their rotational behaviour should be insensitive 
to internal angular momentum transport (see Sills et al. \cite{spt00}). The 
contraction process can be modelled when the evolution of stellar radii is known. 
In case of angular momentum conservation, the rotation period will simply evolve 
with the square of the radius: $P_e = P_i (R_e^2/R_i^2)$. We used the radii from 
the models of Chabrier \& Baraffe (\cite{cb97}), although noting that these radii 
in the VLM regime are poorly verified by observations so far. 
The more recent models of Baraffe et al. (\cite{bcb03}), cover only the mass range 
up to $0.1\,M_{\odot}$, i.e. they end where the mass range of our objects begins, and 
are thus not suitable for our purposes. The radii from Baraffe et al. (\cite{bcb03})
are systematically smaller than in Chabrier \& Baraffe (\cite{cb97}). For 
$M=0.1\,M_{\odot}$, the difference between both models is only 2\% 
at the age of the Pleiades and around 10\% at the age of $\sigma$\,Ori. Thus,
we could underestimate $P_e$ by a factor of $1.1^2/1.02^2 = 1.16$.

\noindent
d) The final ingredient that determines rotational evolution is angular
momentum loss through stellar winds. It has long been known that the rotational
velocity of main sequence stars is proportional to stellar activity, and  
both decline with age following a $t^{-1/2}$ law (Skumanich \cite{s72}), i.e.
the period is proportional to $t^{1/2}$. This simple model has been improved by 
Chaboyer et al. (\cite{cdp95}) to account for the ultrafast rotators in young 
open clusters. This modification introduces a saturation of the activity
at high rotation rates, resulting in $P_e = P_i \exp{(t/\tau_C)}$, where $\tau_C$ 
is the spin-down timescale (see Terndrup et al. \cite{tsp00}, Barnes \cite{b03}). 
Increasing rotation rates for lower mass stars have been successfully modelled using
a mass-dependent saturation level (Barnes \& Sofia \cite{bs96}, Krishnamurthi 
et al. \cite{kpb97}), which scales to the inverse of the global convective
overturn timescale.

In the $\sigma$\,Ori cluster, we measured rotation periods for 23 objects (SE2004). 
Ten of these periods were measured for stars with $0.075<M<0.3\,M_{\odot}$,
the mass regime covered by our Pleiades targets. In the following, we use only
these ten periods as $P_i$. For each of these objects, we determined the radius 
at 3\,Myr (the most probable age for $\sigma$\,Ori, see Zapatero Osorio et al. 
\cite{zbp02}) and 125\,Myr (the age of the Pleiades). For most of these
stars, there is only photometry available, disabling us from determining
the radii empirically. Therefore, the radii were extracted 
from the models of Chabrier \& Baraffe (\cite{cb97}), which are available for 
masses of 0.075, 0.08, 0.09, 0.1, 0.15, 0.2, 0.3$\,M_{\odot}$. For each object, 
we used the model for the mass nearest to the object mass. Thus, we obtained a 
list of $P_i$, $R_i$, and $R_e$, from which we now calculate $P_e$. 

The rotational evolution can then be calculated as
\begin{equation}
P_e = \alpha P_i (R_e^2/R_i^2)
\end{equation}

\begin{figure}[h]
\centering
\resizebox{\hsize}{!}{\includegraphics[angle=-90,width=6.5cm]{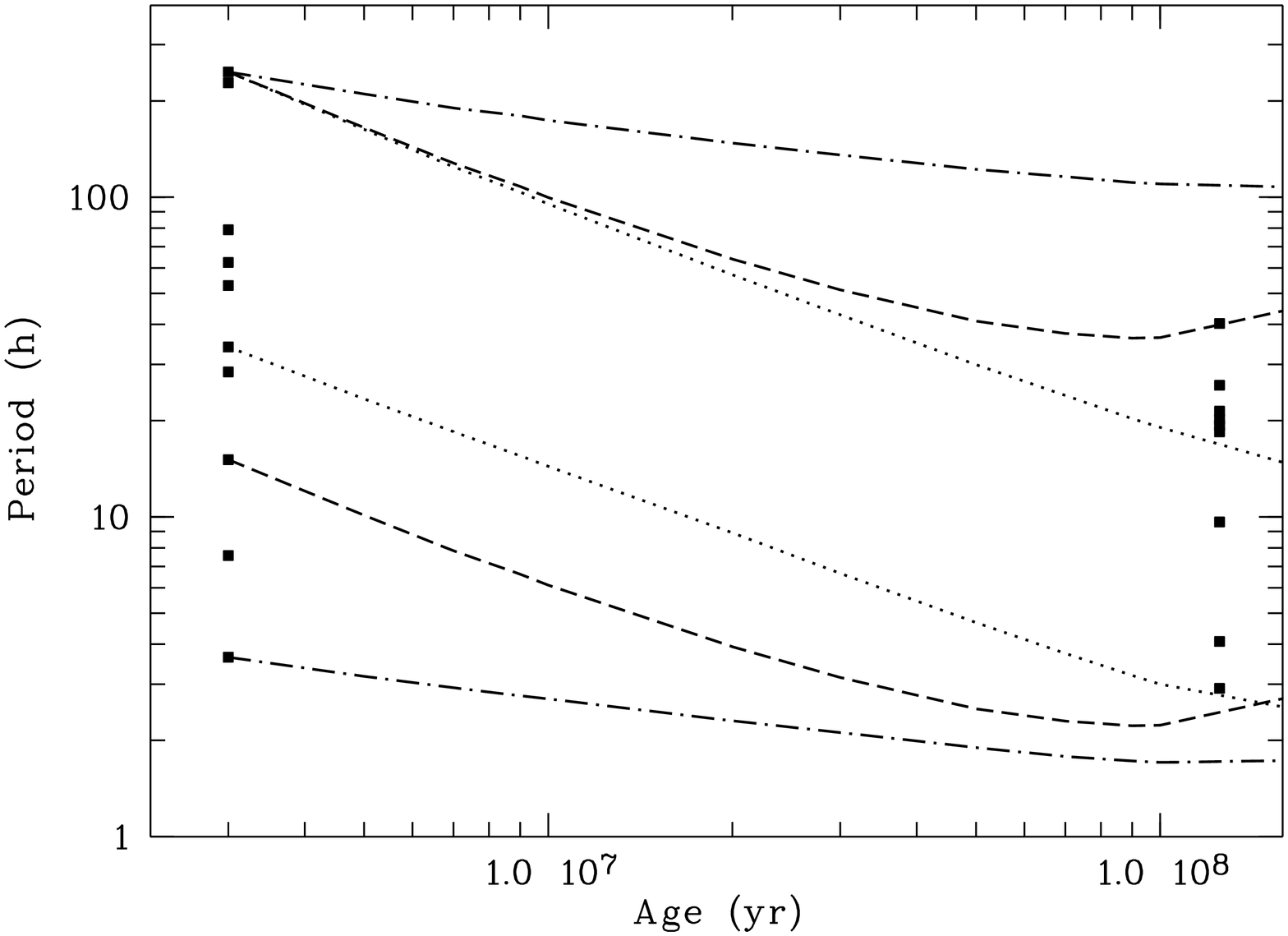}}
\resizebox{\hsize}{!}{\includegraphics[angle=-90,width=6.5cm]{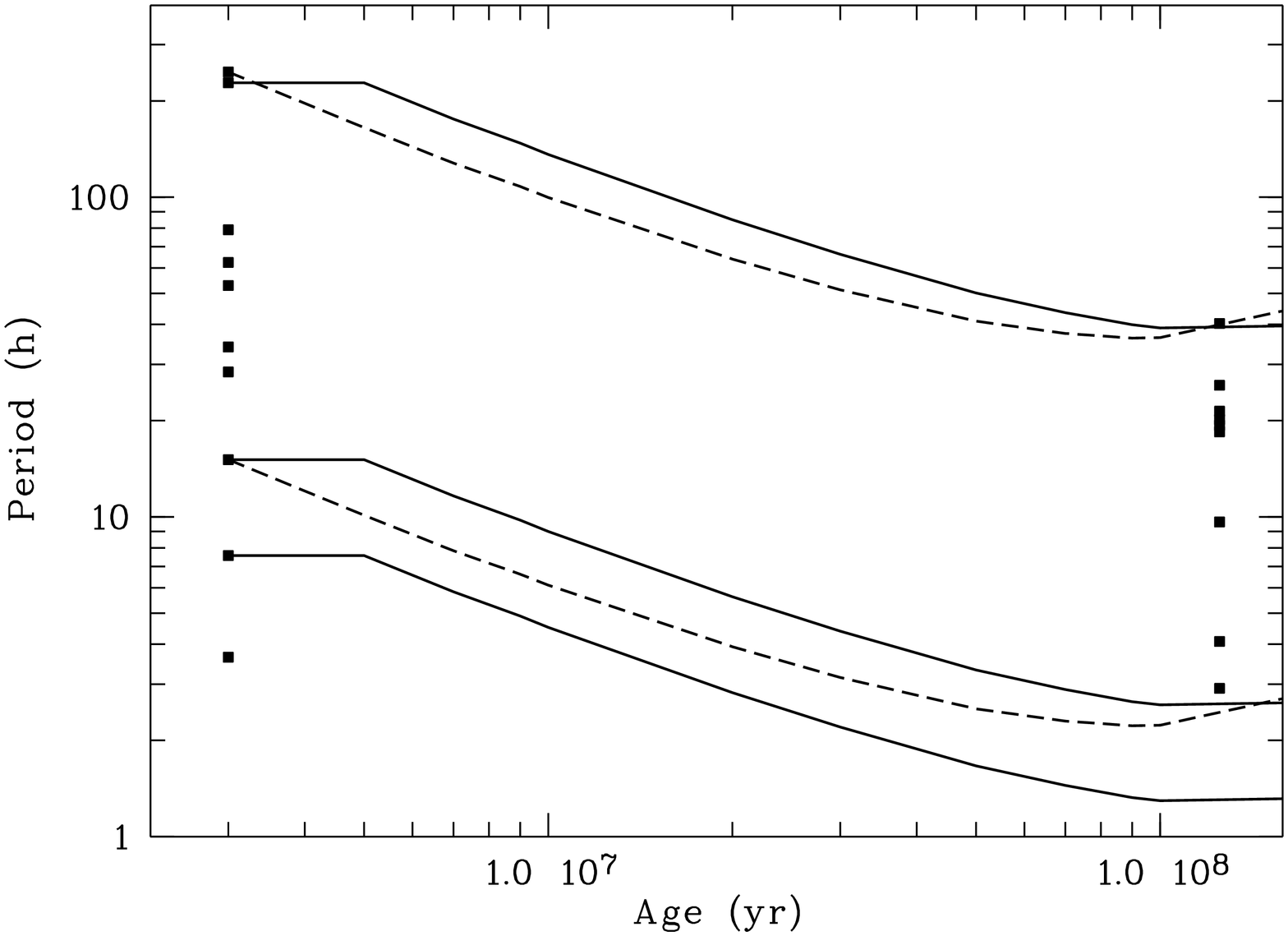}}
\caption{Rotational evolution of VLM objects: The upper panel shows 
the evolution of the rotation period for model A (no braking, dotted lined), 
model B (Skumanich braking, dash-dotted lines), and model C (exponential braking, 
dashed lines). In the lower panel, we compare model C (dashed lines) with
model D (exponential braking with disk-locking, solid lines).}
\label{evo}
\end{figure}

In the following, we consider three different scenarios:
\begin{itemize}
\item{$\alpha = 1$: no rotational braking (model A)}
\item{$\alpha = (t_e - t_i)^{1/2}$: angular momentum loss through stellar
winds following the Skumanich law (model B)}
\item{$\alpha = \exp{(t/\tau_C)}$: angular momentum loss through stellar
winds with saturated activity (model C)}
\item{$\alpha = \exp{(t/\tau_C)}$: model C plus disk-locking (model D)}
\end{itemize}
In Fig. \ref{evo}, we show the rotational evolution for these scenarios.
Model A is shown with dotted, model B with dash-dotted, model C with dashed 
and model D with solid lines. In the following, we discuss the results for each 
of the four models in detail.

{\bf A)} In this model, we simply assume {\it no} angular momentum loss, 
i.e. {\it only} hydrostatic contraction. In this case, the period evolution
is fixed by $P_i$ and the evolution of the radii. For the periods at the
Pleiades age, we obtain $P_e = 0.28 \ldots 17$\,h. In Fig. \ref{evo} 
(upper panel), we show the period evolution for two objects (dotted lines). 
As can be seen from this Figure, half of the $\sigma$\,Ori objects would end 
up with periods below the lower limit of the observed Pleiades periods. Thus, this 
approach is in clear contradiction to the observations in the Pleiades, since 
it produces too high rotation rates for all objects and does neither reproduce 
the upper nor the lower period limit. The lower limit of the known rotation 
periods of VLM objects lies at about 2 to 3\,hours for ages of 3\,Myr (Zapatero 
Osorio et al. \cite{zcb03}), 36\,Myr (Eisl{\"o}ffel \& Scholz \cite{es02}), 
125\,Myr (this work), and $\approx 1$\,Gyr (Clarke et al. \cite{ctc02}), i.e. 
it seems to be nearly constant and independent of age. Since the objects surely 
undergo a significant contraction process, it becomes obvious that there must 
be significant rotational braking. 

{\bf B)} In the second model, we assume a Skumanich law for the rotational
braking. Thus, the period evolution is determined only by the radii of
the objects and the ages of the $\sigma$\,Ori cluster $t_i$ and the Pleiades 
$t_e$. Using $t_i = 3$\,Myr and $t_e = 125$\,Myr (Zapatero Osorio et al. \cite{zbp02},
Stauffer et al. \cite{ssk98}), we obtain periods of $1.7\ldots 110$\,h at 
the age of the Pleiades. Three periods lie outside the limits
defined by the observations in the Pleiades. In Fig. \ref{evo} (upper panel), 
we show the evolution for the slowest and the fastest rotator (dash-dotted
lines). Remarkably, two objects end with periods larger than 100\,h, in clear
contradiction to the observations. Age uncertainties cannot explain this
result, since we would have to choose $>10$\,Myr for the age of $\sigma$\,Ori and 
$<70$\,Myr for the age of the Pleiades to bring both periods below 50\,h. 
According to recent age determinations (Zapatero Osorio et al. \cite{zbp02}, 
Stauffer et al. \cite{ssk98}), this is implausible. Similarly, it is not
possible to explain these long periods with a possible overestimate of the radii,
since this would only result in a decrease of about 16\% (see above). 
Therefore, we can rule out rotational braking following the Skumanich law, even 
for the slowest rotators. This implies that the saturation limit for VLM objects 
lies beyond our maximum observed period of 240\,h in $\sigma$\,Ori, corresponding to 
a rotational velocity $<$5\,kms$^{-1}$. Hence, all our $\sigma$\,Ori objects are in
the saturated regime. This is in agreement with recent studies: From Fig. 5 of Delfosse 
et al. (\cite{dfp98}) and Fig. 10 of Terndrup et al. (\cite{tsp00}), we infer upper 
limits of 3\,kms$^{-1}$ and 6\,kms$^{-1}$ for the saturation threshold.

{\bf C)} In model C, we assume that all stars are beyond the saturation limit.
Thus, the period evolution is determined by the radii of the objects and the
spin-down timescale $\tau_C$. We considered two limiting cases by choosing 
$\tau_C$ either to reproduce the lower or the upper period limit in the Pleiades.
If the model reproduces the lower period limit in the Pleiades of $P=2.9$\,h, we 
obtain $\tau_C = 54$\,Myr. The periods at the age of the Pleiades then range from 
2.9 to 109\,h, and there are four objects with very long periods,
which are not consistent with the available rotational data for the Pleiades,
as already discussed for model B. Thus, it is more plausible to fix
$\tau_C$ by reproducing the upper period limit of the Pleiades. Then, we 
obtain $\tau_C = 150$\,Myr and periods between 0.62\,h and 41\,h. For this scenario,
we show the period evolution for two objects in Fig. \ref{evo} (upper panel, dashed 
lines). Compared with models A and B, model C with $\tau=150$\,Myr clearly delivers the best 
fit to the data. There are, however, still three outliers in the $\sigma$\,Ori period 
sample with very short periods at the age of the Pleiades, which are not consistent with 
the observations. These fast rotators will be discussed in Sect. \ref{fate}.

{\bf D)} This model includes two mechanism for angular momentum loss:
exponential braking through stellar winds as in model C and disk-locking.
For three of the ten $\sigma$\,Ori objects (no. 14, 33, and 80), we have 
strong evidence that they possess an accretion disk. This can be inferred 
from near-infrared colour excess, large amplitude photometric variation,
and accretion indicators in the spectra (see SE2004). For these three
objects, we assume disk-locking up to an age of 5\,Myr. Thus, the rotation
period is constant from 3 to 5\,Myr. This disk-locking scenario was combined
with exponential braking, where the spin-down timescale was determined again
by matching the upper limit of the period distribution in the Pleiades.
In this case, we derived 250\,Myr. In Fig. \ref{evo} 
(lower panel), we show the period evolution for the three objects with disk-locking 
(solid lines), and compare them to model C (dashed lines). Remarkably, two of the 
three outliers in $\sigma$\,Ori from model C show signs of ongoing accretion. 
By including disk-locking for these objects, however, the fit to the observed data 
improves only marginally, as can be seen from Fig. \ref{evo} (lower panel). The 
period evolution for model C and model D is nearly indistinguishable. Thus, from
our periods alone, there is no clear evidence for disk-locking on VLM objects.

Recapitulating, we state that a model with exponential angular momentum loss is able 
to reproduce our observed period distribution. We note that we cannot set definite 
constraints for the spin-down timescale from our data, since $\tau_C$ is extremely 
sensitive to changes in $R_i$ and thus the age of $\sigma$\,Ori. Assuming an age 
of 8\,Myr (the upper limit given by Zapatero Osorio et al. \cite{zbp02}), the values for 
$R_i$ decrease by a factor of 1.5, leading to a increased spin-down timescale of 300\,Myr
(with model C) and 950\,Myr (with model D). Finally, we note that the 
definition of the lower and upper period envelope for the Pleiades is still 
limited by the small number of data points. Therefore, for a precise determination 
of $\tau_C$, better age estimates and more periods are needed. The influence 
of disk-locking could be excluded by using an older period sample as starting 
point.

With all these limitations in mind, we conclude that the spin-down
timescale of VLM objects is most likely a few hundred Myr, and thus clearly
longer than for solar-mass G-type stars (see Barnes \cite{b03}). Previous
estimates of the spin-down timescale in the VLM regime are $\tau_C=246\pm 55$\,Myr 
(Terndrup et al. \cite{tsp00}) and $\tau_C>1$\,Gyr (e.g., Delfosse et al. \cite{dfp98}, 
Barnes \cite{b03}, Sills et al. \cite{spt00}). As noted above, more observational data 
are needed to enable a reliable assessment of $\tau_C$.

Recently, Barnes (\cite{b03}) proposed an appealing interpretation of stellar
rotation periods, where the available period data is interpreted {\it only}
in terms of the magnetic field configuration. According to this model, the 
available periods lie primarly on two sequences, called $I$ and $C$ sequence.
The $I$ sequence produces slowly rotating stars, which exhibit large-scale,
solar-type dynamos. On the other hand, stars on the $C$ sequence are exclusively
fast rotators, because they only possess small-scale, convective magnetic fields.
Due to their low masses, all our targets are fully convective (Chabrier \& Baraffe 
\cite{cb97}). Therefore, all periods should belong to the $C$ 
sequence, for which an exponential spin-down and a lack of slow rotators is 
predicted. Although the model of Barnes (\cite{b03}) probably overestimates 
the spin-down timescales for VLM objects (see above), these predictions are
in nice agreement with our results.

\subsection{The fate of very fast rotators}
\label{fate}

Our periods in the Pleiades can be reproduced from the $\sigma$\,Ori periods
with a model including exponential braking (model C in Sect. \ref{model}). This model, 
however, still produces three $\sigma$\,Ori objects with evolved periods $P_e$ below 
the lower limit of the period distribution in the Pleiades. Two of these fast rotators 
have $P_e<2\,h$. Although these outliers could be discussed away with the field star 
contamination of the $\sigma$\,Ori targets and low number statistics, such an 
interpretation is not satisfying. 

For our $\sigma$\,Ori targets, the critical period $P_{crit}$, at which gravitational 
and centrifugal forces at the stellar surface are balanced (see Porter \cite{p96}), 
lies at $P_{crit} = 3\ldots 7$\,h, where the exact value depends on the mass of the 
object and the exact age of $\sigma$\,Ori. Thus, two of the $\sigma$\,Ori 
objects rotate nearly at breakup velocity. 

The period evolution for these fast rotators (model C in Sect. \ref{model}) was 
compared with the evolution of the breakup period, which is determined by the
evolution of the radii. As the stars contract, their periods decrease with $R^{2}$. 
On the other hand, $P_{crit}$ decreases only with $R^{1.5}$ (Herbst et al. 
\cite{hbm01}). Thus, we find that the fast rotators will probably approach the 
breakup velocity as they get older.

It is not clear what would happen if the rotation period of a star arrives at
its critical period. One extreme scenario would be the complete disruption
of the object after reaching its breakup velocity. In this case, our fast 
rotators in the $\sigma$\,Ori sample would not reach the age of the Pleiades,
and as a consequence the IMF should change somewhat with time.
It seems, however, more probable that the disruption is avoided by
throwing off surface material when approaching the critical velocity. Thus, the 
object might continue to evolve at or near the criticial period. In this scenario, 
the rotational evolution would be influenced by the mass loss, in the sense that an 
additional braking mechanism is involved. Moreover, the oblateness of the object,
caused by its fast rotation, will increase its equatorial radius compared to a 
slowly rotating object. The models of Chabrier \& Baraffe (\cite{cb97}), used for the 
model calculations of Sect. \ref{model}, neglect the influence of rotation on the 
evolution of the radii. For all these reasons, we conclude that the models of Sect. 
\ref{model} are inappropriate for the fastest rotating $\sigma$\,Ori targets.

We note that the oblateness of the fast rotators could induce periodic 
variability, if the object undergoes significant precession. In this case, the 
size of the visible surface is modulated by the rotation period. Therefore, it 
might be that the periodic variability of very fast rotators in the $\sigma$\,Ori
cluster is not caused by co-rotating spots on the surface, but by the oblateness
and precession of the object.

\section{Conclusions}
\label{conc}

We report a photometric monitoring campaign for VLM stars in the
Pleiades. From lightcurve analysis, we derived rotation periods for nine
Pleiades members with masses between 0.08 and 0.25$\,M_{\odot}$. Their periodic
variability is likely caused by magnetically induced spots rather than 
inhomogenuously distributed dust clouds, since the targets are still too hot for 
dust condensation, but probably hot enough for magnetic surface activity. The 
lightcurves show very low amplitudes compared with more massive Pleiades
stars, indicating that either the relative spotted area, the asymmetry of the 
spot distribution or the intensity contrast between spots and photosphere is 
reduced in the VLM regime. From our lightcurves, we see clear evidence for the 
temporal evolution of the spot patterns on timescales of about two weeks.

The rotation periods range from 2.9\,h to 40\,h, although our time series
analysis is sensitive to periods up to 300\,h. Comparing with the known
periods for solar-mass stars in the Pleiades, we find a clear lack of 
slow rotators among VLM stars, in agreement with previous $v\sin i$ 
studies. In the VLM regime, the periods tend to decrease towards lower
masses. Since this correlation has already been found for very young VLM
objects, it must have its origin in the earliest phases of their evolution.

By combining the previously published periods for the young $\sigma$\,Ori
cluster (age 3\,Myr, SE2004) with the periods in the Pleiades, we studied the
rotational evolution in the VLM regime. Since the lower period limit
is nearly constant at all ages, despite of the hydrostatic contraction process
of the objects, there must be significant angular momentum loss. It was found 
that a Skumanich type angular momentum loss law ($P\propto t^{1/2}$) is not 
applicable in the VLM regime. Instead, the period evolution can be understood with
saturated angular momentum loss following $P \propto \exp{(t)}$. 
We see no significant evidence for a contribution of rotational braking
through star-disk interaction. Our best-fitting model cannot account for 
the fastest rotators in the $\sigma$\,Ori sample. These objects 
rotate nearly at breakup velocity, and they will probably approach their breakup 
velocity as they get older. Therefore, their rotational evolution could be 
influenced strongly by mass loss and oblateness.

The observed lack of slow rotators and the exponential period evolution may
be understood as a consequence of a convective magnetic field, as described by
Barnes (\cite{b03}). All our targets are convective throughout their evolution, 
and thus unable to sustain a solar-type large-scale dynamo, which is the origin of 
the Skumanich type angular momentum loss law. Instead, VLM objects may exhibit 
small-scale fields, likely of turbulent nature, implying high rotation rates and 
an exponential spin-down, as observed. In addition, the turbulent dynamo scenario 
predicts rather symmetric spot distributions and weak surface activity, confirmed 
by the low lightcurve amplitudes.

\begin{acknowledgements}
      This paper greatly benefits from the application of the difference 
      imaging technique. It is therefore a pleasure to acknowledge the
      cooperation with the WeCAPP team, who delivered the software for 
      this technique. In particular, we thank Arno Riffeser for valuable
      suggestions for the data reduction process. An implementation
      of the CLEAN algorithm was kindly provided by David H. Roberts.
      We thank the Calar Alto staff for generous support during the 
      observing run. The comments of an anonymous referee helped us to 
      significantly improve this paper. The Open Cluster Database, as 
      provided by C.F. Prosser (deceased)  and J.R. Stauffer, may be 
      accessed at http://cfa-ftp.harvard.edu/~stauffer/, or by anonymous 
      ftp to cfa-ftp.harvard.edu, cd /pub/stauffer/clusters/. This work was 
      supported by the German \emph{Deut\-sche For\-schungs\-ge\-mein\-schaft, DFG\/} 
      grants Ei~409/11--1 and HA3279/2--1. 
\end{acknowledgements}

\appendix

\section{Period search in phase space}

We investigated as alternatives the period search algorithms of Dworetsky (\cite{d83}) and
Cincotta et al. (\cite{cmn95}) by applying them to the lightcurve of the object 
BPL129. For this object, our analysis (Sect. \ref{per}) reveals a convincing 
periodicity with $P=9.64$\,h (see also Fig. \ref{phase}). Any alternative period 
search method should also be able to recover this period.

In contrast to the frequency space based Fourier techniques, the methods of 
Dworetsky (\cite{d83}) and Cincotta et al. (\cite{cmn95}) work in phase space. They 
arrange the data points according to their phase for a number of test periods. For each period, 
they compute a statistical parameter, which allows to examine whether the time series 
contains this period. In the method of Dworetsky (\cite{d83}), this 
parameter is the 'string length', the sum of the distances of the consecutive data points 
in phase space. If the time series contains a periodicity, the string length should show 
a minimum at this period. The algorithm of Cincotta et al. (\cite{cmn95}) relies
on the entropy minimization of the lightcurve in phase space. Test parameter is the
entropy, which should be normally distributed if no period is present 
(Cincotta et al. \cite{chm99}).

\begin{figure}[htbd]
\centering
\resizebox{\hsize}{!}{\includegraphics[angle=-90,width=6.5cm]{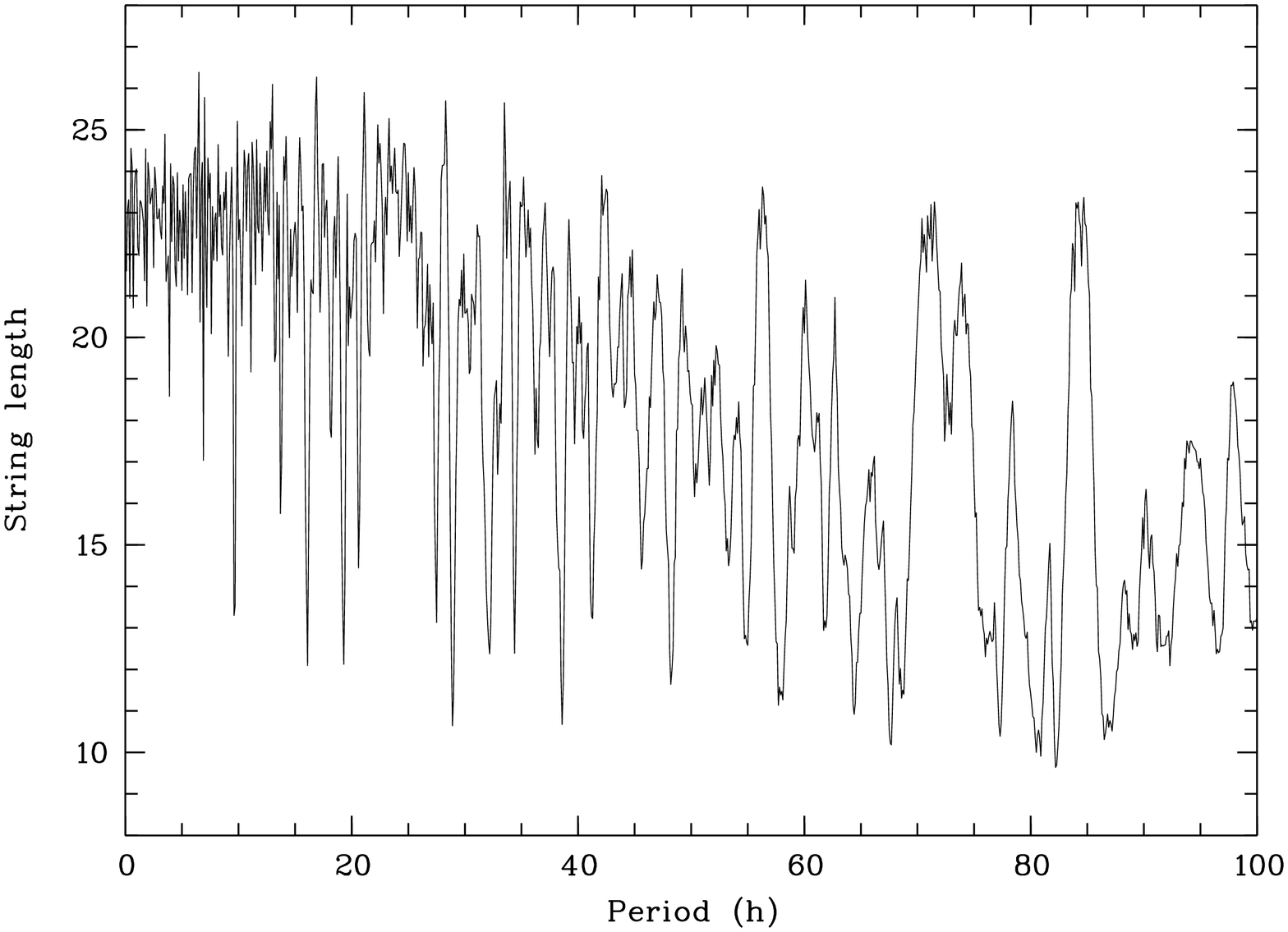}}
\resizebox{\hsize}{!}{\includegraphics[angle=-90,width=6.5cm]{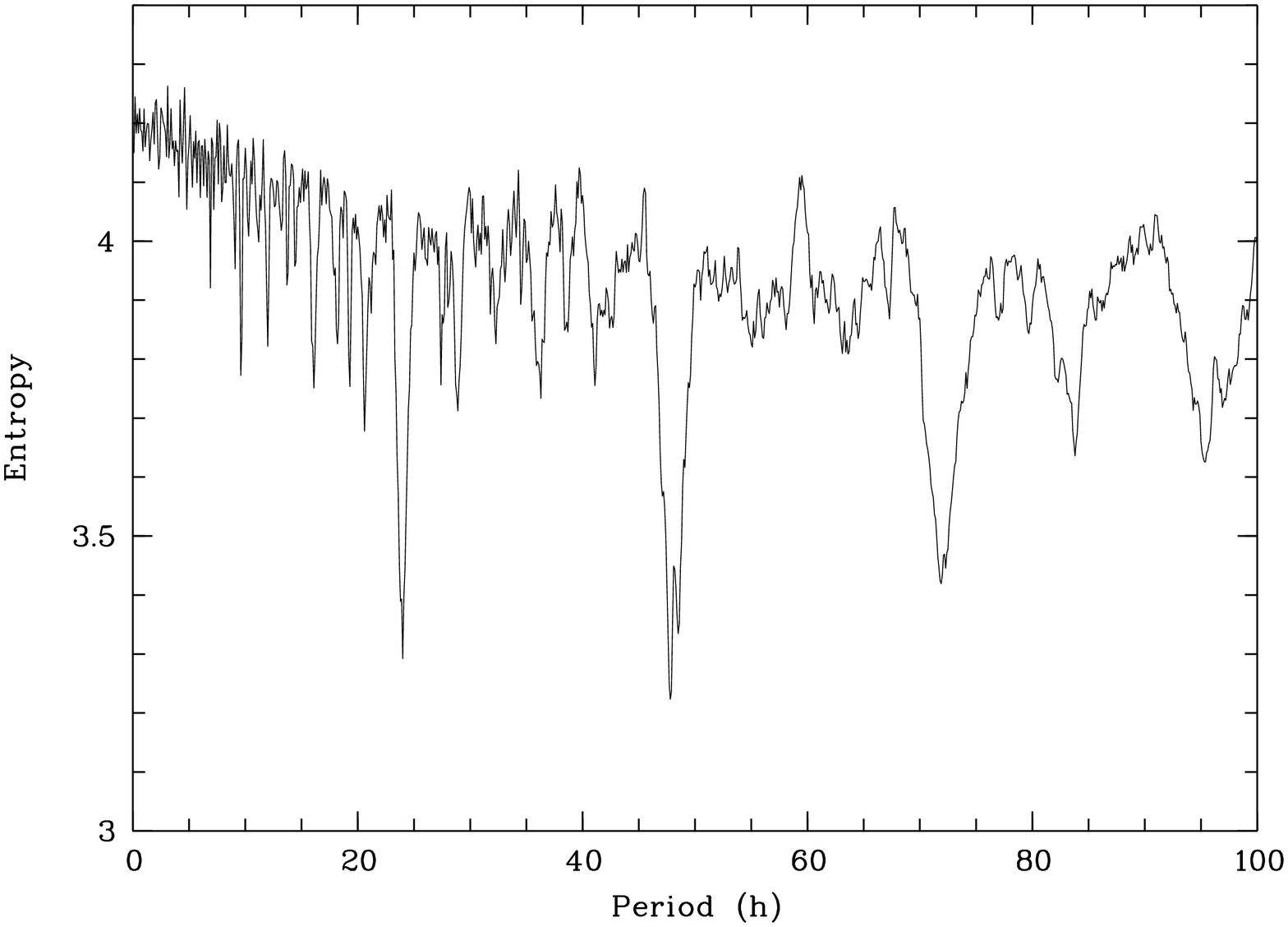}}
\caption{String length (upper panel) and entropy (lower panel) vs. period for
the lightcurve of BPL129 (see text for explanation).}
\label{phasespace}
\end{figure}

We computed string length and entropy for the lightcurve of BPL129.
In Fig. \ref{phasespace}, we show the results. Both plots show a
local minimum at 9.6\,h, the adopted period from our time series analysis. 
In both plots, however, there are many other lower minima. The string length
distribution is very noisy, there exist many minima with similar 
depth. The global minimum lies at $P=82.1$\,h, but the corresponding phase plot 
shows no periodicity at all. The reason for the failure of the algorithm is 
probably the large number of data points. Dworetsky (\cite{d83}) argues that 
his method is optimum for very few data points and random sampling. In the 
literature, the string length method is applied to lightcurves with typically 
less than 20 data points (e.g., Bouvier et al. \cite{bcf93}, Joergens et al. 
\cite{jgn01}). Our results suggest that for many data points it is much more 
reliable to use Fourier techniques. 

On the other hand, the entropy shows deep minima at multiples of one day, but
the corresponding phase plots show no periodicity. From Fig. \ref{sampling}, it 
is obvious that for a period of one day all data points will have phases between
0.0 and 0.2 or between 0.8 and 1.0, leading to a minimized entropy. Hence, 
these minima are artifacts caused by the regular gaps between the observing 
nights. The method is therefore probably not useful for datasets with regular
gaps.

Recapitulating, we found that both phase spaced techniques are not applicable
to our data. For lightcurves with many data points and clumped data point
distribution, Fourier techniques in combination with plausibility checks, 
as outlined in Sect. \ref{per}, are clearly superior and probably the best 
way to search for periods.

\end{document}